\documentclass[a4paper]{article}

\usepackage{jheppub} 
                     
\usepackage{graphicx,color}
 \usepackage{bm}
   \usepackage{amsmath}
    \usepackage{amssymb}    
     \usepackage{pifont}
 \usepackage{stmaryrd}
\usepackage{standalone}
\usepackage{slashed}
\usepackage{multirow}
\usepackage{makecell}
\usepackage{tabu}
\usepackage{hhline}
\usepackage{colortbl}

\newcommand*{\FigPath}{./Figs}%
\usepackage{bbm}
\usepackage{orcidlink}

\newcommand{\xbj}{\ensuremath{x}}
\newcommand{\zh}{\ensuremath{z_{h}}}

\usepackage{subfig}

\DeclareRobustCommand{\eq}[1]{Eq.~\eqref{eq:#1}}
\DeclareRobustCommand{\eqs}[2]{Eqs.~\eqref{eq:#1} and \eqref{eq:#2}}
\DeclareRobustCommand{\fig}[1]{Fig.~\ref{fig:#1}}

\usepackage{tikz}

\newcommand{\bea}{\begin{eqnarray}}
\newcommand{\eea}{\end{eqnarray}}

\newcommand{\nn}{\nonumber\\}
\newcommand{\be}{\begin{equation}}
\newcommand{\ee}{\end{equation}}
\newcommand{\ba}{\begin{eqnarray}}
\newcommand{\ea}{\end{eqnarray}}
\newcommand{\la}{\langle}
\newcommand{\ra}{\rangle}

\newcommand{\half}{{1\over 2}}

\def\bfkperp{{\bm k}_T}

\def\bfhp{\hat{\bm h}}
\def\bfPhperp{{\bm P}_{h\perp}}
\def\Phperp{P_{h\perp}}

\newcommand{\df}{\mathrm{d}}

\DeclareRobustCommand{\eq}[1]{Eq.~\eqref{eq:#1}}
\DeclareRobustCommand{\eqs}[2]{Eqs.~\eqref{eq:#1} and \eqref{eq:#2}}
\DeclareRobustCommand{\sec}[1]{Sec.~\ref{sec:#1}}

\def\avkperp{\la k_\perp^2 \ra}
\def\avpperp{\la p_\perp^2 \ra}

\newcommand{\mh}{ m_h }

\def\qT{\bm q_T^{ } }

\newcommand{\bsi}{{\bar \psi}}
\newcommand{\slk}{{\slashed{k}}}
\newcommand{\slp}{{\slashed{p}}}

\newcommand{\slA}{{\slashed{A}}}

\newcommand{\sln}{\slashed{n}}

\newcommand{\barD}{{\bar D}}
\newcommand{\barE}{{\bar E}}
\newcommand{\barF}{{\bar F}}
\newcommand{\barG}{{\bar G}} 
\newcommand{\barH}{{\bar H}}  

\newcommand{\barf}{{\bar f}}

\newcommand{\barh}{{\bar h}}
\newcommand{\beq}{\begin{equation}} 
\newcommand{\eeq}{\end{equation}} 
\newcommand{\bega}{\begin{eqnarray}} 
\newcommand{\ega}{\end{eqnarray}} 
\newcommand{\calW}{{\cal W}} 

\newcommand{\tilq}{{\tilde q}}
\newcommand{\calw}{{\cal W}} 
\newcommand{\breq}{\breve {q}} 

\def\kT{\bm k_{T}}
\def\bT{\bm b_{T}}

\usepackage[normalem]{ulem}

\title{Next-to-next-to-leading power corrections to unpolarized Semi-Inclusive Deep Inelastic Scattering}

\newcommand*{\ODU}{Department of Physics, Old Dominion University, Norfolk, VA 23529, USA}
\newcommand*{\PSU}{Division of Science, Penn State University Berks, Reading, Pennsylvania 19610, USA}
\newcommand*{\JLAB}{Jefferson Lab, Newport News, VA 23606, USA}

\author{Ian~Balitsky$^{\orcidlink{0009-0005-5170-6518},a,c}$}  
\author{Alexei~Prokudin$^{\orcidlink{0000-0001-5956-4159},b,c}$}
\affiliation{$^a$\ODU}
\affiliation{$^b$\PSU}
\affiliation{$^c$\JLAB}

\emailAdd{balitsky@jlab.org}
\emailAdd{prokudin@jlab.org}

\preprint{JLAB-THY-26-4591}

\abstract{
Semi-Inclusive Deep Inelastic Scattering (SIDIS) is a key tool for exploring the three-dimensional structure of the nucleon through Transverse Momentum Dependent parton distributions and fragmentation functions. While leading-power contributions to the SIDIS cross-section are well established, next-to-leading power (NLP) corrections of order $1/Q$ and next-to-next-to-leading power (NNLP) corrections of order $1/Q^2$ to the hadronic tensor have only recently begun to be systematically investigated. These corrections are essential for reliable phenomenology and interpretation of modern high-precision data.
In recent papers by one of the authors, NNLP corrections to the Drell–Yan process were derived using the rapidity factorization formalism. In the present work, we extend this approach to SIDIS and obtain analytic expressions for the unpolarized structure functions. We derive NNLP corrections that include convolutions of unpolarized distributions, $f_1$, with unpolarized fragmentation functions, $D_1$, and Boer–Mulders functions, $h_1^\perp$, with Collins fragmentation functions, $H_1^\perp$. We compare our results with previous formulations, provide numerical studies, confront our predictions with HERMES and COMPASS measurements, and present predictions for future experiments at Jefferson Lab and the Electron-Ion Collider.
}


\begin{document} 
\allowdisplaybreaks
\maketitle 

\section{Introduction}
Semi-Inclusive Deep Inelastic Scattering is one of the most studied processes at various experimental facilities, such as HERMES~\cite{HERMES:2004vsf} (DESY), COMPASS~\cite{COMPASS:2007rjf} (CERN), Jefferson Lab~\cite{Dudek:2012vr}, and the future Electron-Ion Collider~\cite{AbdulKhalek:2021gbh}. In this process, a lepton with momentum $l$ scatters off a nucleon or a nucleus with momentum $P$, and the target is destroyed, producing hadrons, one of which is detected with momentum $P_h$, alongside the scattered lepton that has momentum $l'$. When the transverse momentum of the produced hadron is comparable to the virtuality of the exchanged photon, $Q^2 = - q^2 = -(l-l')^2$, or the transverse momentum $\Phperp$ is integrated, the collinear QCD factorization~\cite{Collins:2011zzd} is applicable, and one describes the process with collinear parton distribution and fragmentation functions that depend on the Bjorken $x$ and $\zh$, respectively. When the transverse momentum is small enough, $\Phperp/\zh \ll Q$, the Transverse Momentum Dependent factorization is valid. The cross section is described in terms of the Transverse Momentum Dependent distribution and fragmentation functions~\cite{Boussarie:2023izj}, collectively called TMDs, which encode the three-dimensional (3D) structure of the nucleon.

In recent years, the 3D nucleon structure has attracted a lot of interest in the experimental and theoretical communities; see Ref.~\cite{Boussarie:2023izj} for a review. The SIDIS cross section is proportional to the convolution of the leptonic and hadronic tensors. In pioneering works by Mulders and Tangerman~\cite{Mulders:1995dh} and Kotzinian~\cite{Kotzinian:1994dv}, the leading power and next-to-leading power $\sim 1/Q$ were calculated for the hadronic tensor, which allowed a clear interpretation of the experimental data and led to successful phenomenological studies of both polarized and unpolarized SIDIS. These pioneering works were followed by a thorough investigation~\cite{Collins:2011zzd} of the factorization theorems that led to a good understanding of the scale dependence of TMDs. Recently, phenomenology of TMDs has been performed with TMD evolution~\cite{Collins:2011zzd}. We refer the reader to Chapter 10 of the TMD handbook~\cite{Boussarie:2023izj} for further introduction to the sub-leading contributions in SIDIS and Drell–Yan.

Until a few years ago, the next-to-next-to-leading power corrections $\sim 1/Q^2$ to the hadronic tensor of SIDIS were unknown. These corrections are potentially very important given the high precision of the existing and future experimental data. Moreover, some of the structure functions, for example $F_{UU,L}$, that encode the longitudinally polarized virtual photon, were neglected as they include neither leading nor next-to-leading power contributions. The structure function $F_{UU,L}$ is studied~\cite{HallCSIDIS:2025plr} at Jefferson Lab, and therefore it is timely to investigate the next-to-next-to-leading contributions in SIDIS. In addition, NNLP corrections can help to better understand the 3D structure of the nucleon. In unpolarized scattering, they may be important for the understanding of the observed multiplicities, $\cos \phi_h$~\footnote{$\phi_h$ is the azimuthal angle of the produced hadron with respect to the lepton scattering plane.}, and $\cos 2\phi_h$ modulations of the cross sections. This, in turn, will be important for the understanding of spin asymmetries and the extraction of polarized TMDs, for instance the transversity function~\cite{Ralston:1979ys}, which is the only source of direct information about the tensor charge of the nucleon.

Recently, NLP corrections were investigated by Ebert, Gao, and Stewart in Ref.~\cite{Ebert:2021jhy}, by Gamberg et al. in Ref.~\cite{Gamberg:2022lju}, and by Rodini and Vladimirov in Ref.~\cite{Rodini:2023plb}. The TMD operator product expansion at NLP was studied by Vladimirov, Moos, and Scimemi in Ref.~\cite{Vladimirov:2021hdn}. Kinematic power corrections to TMD factorization were studied by Vladimirov in Ref.~\cite{Vladimirov:2023aot}, and by Pilo~neta and Vladimirov in Ref.~\cite{Piloneta:2025jjb}, and kinematic next-to-next-to-leading power corrections were derived for the hadronic tensor in SIDIS. Jet production in SIDIS at next-to-leading power was investigated in Ref.~\cite{Jaarsma:2025ksf}.

In this paper, we will extend the rapidity factorization formalism~\cite{Balitsky:2017flc} of one of the authors, which was used for Drell–Yan in Refs.~\cite{Balitsky:2020jzt,Balitsky:2024ozy}, and apply it to Semi-Inclusive Deep Inelastic Scattering. We will derive $\sim 1/Q^2$ corrections to the hadronic tensor that include convolutions of unpolarized distributions, $f_1$, with unpolarized fragmentation functions, $D_1$, and Boer–Mulders functions~\cite{Boer:1997nt}, $h_1^\perp$, with Collins fragmentation functions~\cite{Collins:1992kk}, $H_1^\perp$, and obtain formulas for the unpolarized structure functions in Semi-Inclusive Deep Inelastic Scattering: $F_{UU,T}$, $F_{UU,L}$, $F_{UU}^{\cos\phi_h}$, and $F_{UU}^{\cos2\phi_h}$. We will compare our results with the existing formulas from~\cite{Ebert:2021jhy,Piloneta:2025jjb} and other papers, and present numerical estimates and comparisons with the experimental data from HERMES and COMPASS. We will also provide estimates for future measurements of the ratio $F_{UU,L}/F_{UU,T}$ at Jefferson Lab and the future Electron-Ion Collider.

The paper is organized as follows: in \sec{kinematics}, we will introduce kinematics and our notations for SIDIS; in \sec{tensor}, we will extend the rapidity factorization formalism~\cite{Balitsky:2024ozy} to SIDIS and derive the SIDIS hadronic tensor including next-to-next-to-leading power corrections $\sim 1/Q^2$. In \sec{structure}, we will derive expressions for the structure functions $F_{UU,T}$, $F_{UU,L}$, $F_{UU}^{\cos\phi_h}$, and $F_{UU}^{\cos2\phi_h}$. In \sec{numerics}, we will provide numerical estimates, compare them with the experimental data, and compare our formulas with the existing results. In particular, we will investigate $F_{UU,T}$ due to transversely polarized photons in \sec{fuut} and compare our results with the experimental data. We will explore $F_{UU,L}$ due to the longitudinal photon polarization in \sec{fuut} and give predictions for $R_{\rm SIDIS}$ to be measured at Jefferson Lab and the EIC. In \sec{fuucosphi}, we will study the subleading $\sim 1/Q$ structure function $F_{UU}^{\cos\phi_h}$ and the corresponding asymmetry, and in \sec{fuucos2phi}, we will investigate the structure function $F_{UU}^{\cos2\phi_h}$, which has the leading contribution from the convolution of Boer–Mulders~\cite{Boer:1997nt} and Collins~\cite{Collins:1992kk} functions. We conclude and discuss future directions in \sec{conclusions}.

\section{Semi-Inclusive Deep Inelastic Scattering process}
\label{sec:sidis}
\subsection{Kinematics}
\label{sec:kinematics}

\begin{figure}[htb]
\centering
	\includegraphics[width=7.2cm]{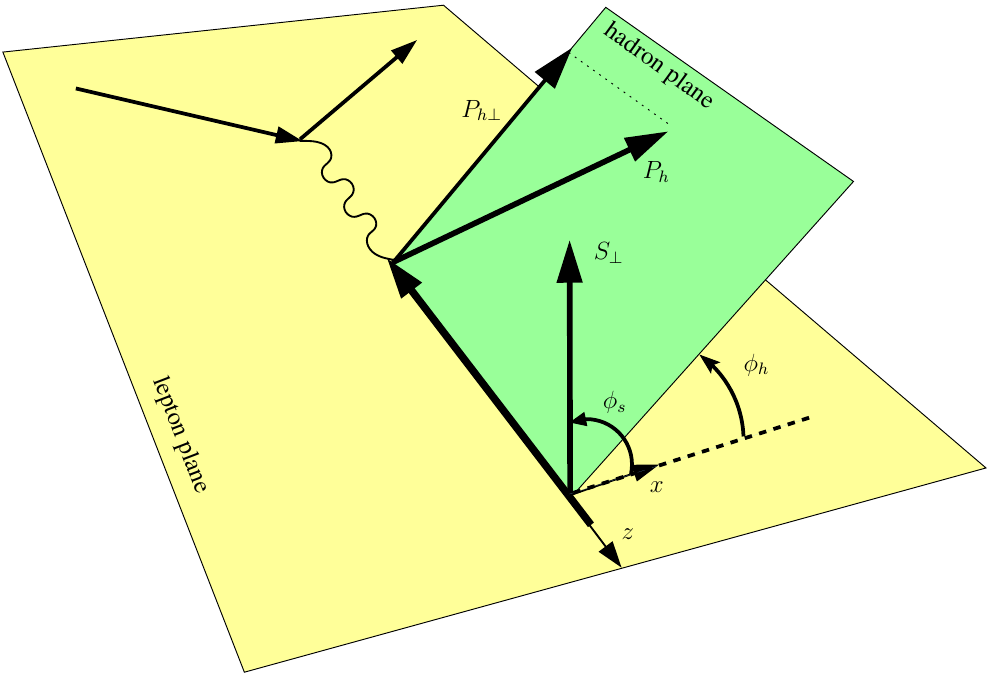}
        \caption{\label{fig-kin-SIDIS}
    	Kinematics of the SIDIS process $lp\to l^\prime h X$
	in the one photon exchange approximation in the Trento frame~\cite{Bacchetta:2004jz}.}
\vspace{5mm}
\end{figure}

In this section we will define the kinematics of the process and introduce our notations. The Semi-Inclusive Deep Inelastic Scattering process (SIDIS)  
\begin{align}
\ell(l)+p(P) \to \ell(l') +h(P_h) +  X
\,,
\end{align}
in the single-photon exchange approximation is sketched in
Fig.~\ref{fig-kin-SIDIS}. Here, $l$ and $P$ are the momenta of the incoming
lepton $\ell$ and the nucleon $p$; $l^\prime$ and $P_h$ are the momenta of the outgoing
lepton $\ell$ and of the detected produced hadron $h$. The center-of-mass energy of the process is $s=(l+P)^2$, the virtual-photon momentum $q=l-l^\prime$ 
defines the $z$-axis of the Trento $\gamma^* P$ frame~\cite{Bacchetta:2004jz}, its virtuality is $Q^2=-q^2$. Vectors $l^\prime$ and $l$ define the lepton plane and $l^\prime$ points in the direction of the $x$-axis
from which azimuthal angles are counted. The produced hadron $h$ has momentum $P_h$ and its transverse momentum is $\Phperp$. Vectors $q$ and  $P_h$ define the hadron plane. Fully differential cross-section at $Q^2 \ll M_Z^2$ reads~\cite{Kotzinian:1994dv,Mulders:1995dh,Bacchetta:2006tn}:
\begin{align}
\df \sigma = \frac{\alpha_{em}^2}{(l\cdot P) Q^4} L_{\mu\nu}W^{\mu\nu} \frac{\df^3 l'}{2 E_{l'}} \frac{\df^3 P}{2 E_P}
\end{align}
where $L_{\mu\nu}$ is the leptonic tensor
\begin{align}
L_{\mu\nu} = 2(l_\mu l'_\nu + l'_\mu l_\nu  - (l\cdot l') g_{\mu\nu}) + 2 i \lambda_l\epsilon^{\mu\nu\rho\sigma}l_\rho l'_\sigma \; ,
\end{align}
where $\lambda_l$ is the helicity of the lepton, and $W_{\mu\nu}$ is the hadronic tensor~\footnote{Here $\sum_X$ denotes the sum over full set of ``out''  states and 
$J_\mu=\sum e_f\bsi^f\gamma_\mu\psi_f$ is the electromagnetic current. We take into account only $u,d,s$ quarks and consider them massless. }
\begin{eqnarray}
\hspace{-1mm}
W_{\mu\nu}(q)~&\stackrel{\rm def}{=}&~\frac{1}{(2\pi)^4}\sum_X\!\int\! \df^4x~e^{iq\cdot x}
\langle P|J_\mu(x)|P_h+X\rangle\langle P_h+X|J_\nu(0) |P\rangle \; .
\label{W}
\end{eqnarray}
The relevant kinematical variables expressed via Lorentz invariants
are:
\begin{align}
   x  = \frac{Q^2}{2\,P\cdot  q}, \;\;
   y = \frac{P \cdot  q}{P \cdot  l}, \;\;
   \zh = \frac{P \cdot  P_h}{P\cdot  q}.\;\;
\label{eq:xyz}\;\;\;\;
\end{align}
In addition to $x$, $y$, and $z_h$, the cross section is also differential
in the azimuthal angle $\phi_h$ of the produced hadron, in the square
of the hadron's momentum component $\Phperp$, and in the azimuthal angle of the outgoing lepton $l'$ around beam axis, $\psi\approx \phi_S$ in $Q\gg M$ limit.  

In the one photon exchange approximation SIDIS process can be  characterized by 18 independent structure functions~\cite{Kotzinian:1994dv,Mulders:1995dh,Bacchetta:2006tn}.
We will consider unpolarized scattering where 4 structure functions contribute to the process~\cite{Kotzinian:1994dv,Mulders:1995dh,Bacchetta:2006tn}:
\begin{align} \label{eq:SIDIS-leading}
 \frac{\df\sigma}{\df\xbj \, \df y \, \df \zh \, \df\phi_h \, \df\Phperp^2} &
 =  \, \sigma_0 \biggl[
   F_{UU,T} + p_1 F_{UU,L}+ 
          \cos(\phi_h)\,p_3 \, F_{UU}^{\cos \phi_h}+\cos(2\phi_h)\,  
   p_1\,F_{UU}^{\cos 2\phi_h}
 \biggr]
\,,\end{align}
where
\begin{align}
\sigma_0 = \frac{2 \pi \alpha_{\rm em}^2}{\xbj\,y\,Q^2} \biggl(1-y+\frac12y^2\biggr)
\end{align}
Up to corrections suppressed as $1/Q^2$, the kinematic prefactors $p_i$ in \eq{SIDIS-leading} are given by~\cite{Bastami:2018xqd}
\begin{align} \label{eq:y-prefactors}
 p_1 &= \frac{1-y}{1-y+\frac12\,y^2}
\,,\quad
 p_3 = \frac{(2-y)\sqrt{1-y}}{1-y+\frac12\,y^2}
\,.\end{align}

The structure functions in \eq{SIDIS-leading} implicitly depend on $\xbj$, $\zh$, $\Phperp^2$ and $Q^2$.
Their superscripts indicate the azimuthal dependence. The subscripts encode the beam  and target polarizations.
The first subscript $UU$ denotes the unpolarized beam and the target, the second superscript refers to either longitudinal, $L$, or transverse, $T$, virtual photon polarizations.

\begin{figure*}[h]
 \centering
\subfloat[photon-hadron frame]{ 
\includegraphics[width=0.45\textwidth]{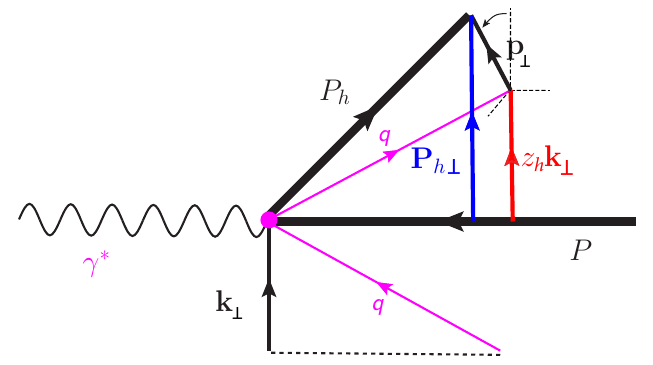} 
}
 \hspace{1cm}
\subfloat[hadron-hadron frame]{ 
\includegraphics[width=0.45\textwidth]{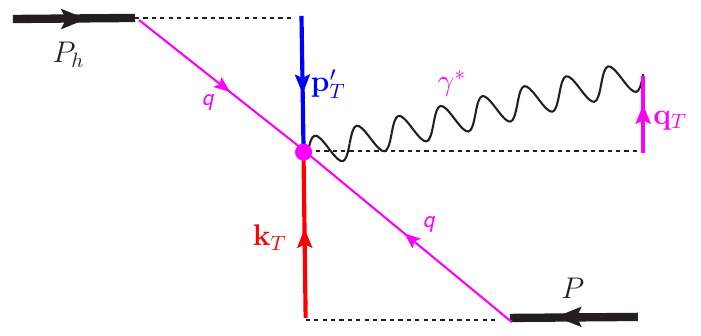} 
}
 \caption{Illustration of the two frames used to describe the kinematics of the SIDIS process, as discussed in the text. The figure is from Ref.~\cite{Boussarie:2023izj}.}
 \label{fig:SIDIS_frames}
\end{figure*}

For the SIDIS process there are two distinct choices~\cite{Collins:2011zzd,Boussarie:2023izj} for frames, see Fig.~\ref{fig:SIDIS_frames}. In the hadron-hadron frame, Fig.~\ref{fig:SIDIS_frames} (b),  one defines transverse direction relative to the incoming hadron $p$ and produced hadron $h$, each of those having zero transverse momentum. In this frame all transverse directions will be labeled with a subscript ``$\bm T$". In this frame the transverse momentum of the virtual photon $\qT$ is non zero and is related to the transverse momentum $\bm k_T$ of the parton in $p$ and the fragmenting parton $\bm p'_T$ relative to the detected hadron $h$.

Their relation in the region of validity of the TMD factorization theorem,  ${\bm q_T^2}/ Q^2 \ll 1$, reads 
\begin{align}
    \qT = - {\bm k_T} + {\bm p'_T} 
\end{align}
The second choice is the photon-hadron frame~\cite{Bacchetta:2004jz,Boussarie:2023izj}, Fig.~\ref{fig:SIDIS_frames} (a). In this case $\bm q$ is aligned with $z$ axis and the incoming hadron still has the vanishing transverse momentum. In this frame all transverse directions will be labeled with a subscript ``$\bm \perp$". The interpretation of the transverse momentum of the parton in $p$ is the same,  ${\bm k_\perp}$, however the outgoing hadron $h$ has transverse momentum ${\bm p_\perp}$ relative to the fragmenting parton and the transverse momentum of the produced hadron $\Phperp$ in this frame is 
\begin{align}
    \bfPhperp = z_h{\bm k_\perp} + {\bm p_\perp} 
\end{align}
For a sufficiently large $Q^2\gg M^2$ the two frames are related as follows
\begin{align}
   {\bm k_\perp} = {\bm k_T}   \; ,\; {\bm p_\perp} = -z_h {\bm p'_T}   \; ,\; \bfPhperp  = -z_h \qT\; .
   \label{eq:approx}
\end{align}
The subtleties of the two frames and the definitions of the convolutions are discussed in Ref.~\cite{Piloneta:2025jjb}. In order to simplify our discussion we will work in the approximation $Q^2 \gg M^2$ where Eqs.~(\ref{eq:approx}) hold.

Structure functions in \eq{SIDIS-leading} are described in terms of convolutions of TMDs PDFs, $f$ and TMD FFs, $D$, in the region where the TMD factorization is valid, and the generic structure of the convolution in our approximation reads~\cite{Mulders:1995dh,Bacchetta:2006tn}
\begin{align} 
 {\cal C}\left[\omega\;f\;D\right] &
 = x \sum_a  H_{aa}(Q^2,\mu^2)\int \df^2 {\bm k_\perp} \, \df^2{\bm p_\perp}
 	\; \delta^{(2)}(\zh {\bm k_\perp} +  {\bm p_\perp}-\bfPhperp) \omega \,f^{a}(x,{\bm k_\perp^2})\ D^{a}(\zh,{\bm p_\perp^2})
 \nonumber\\
&= x \sum_a H_{aa}(Q^2,\mu^2)\int \df^2 {\bm k_\perp} \, 
   \omega \,f^{a}(x,{\bm k_\perp^2})\ D^{a}(\zh,(\bfPhperp - \zh {\bm k_\perp})^2) \label{eq:def-convolution-integral}
 \\   
& = x \sum_a H_{aa}(Q^2,\mu^2)\int \df^2 \kT \, \df^2{\bm p'_T}
 	\; \delta^{(2)}(\kT -  {\bm p_T'} +\qT)
 \omega \,f^{a}(x,{\bm k_T^2})\ D^{a}(\zh,(-\zh {\bm p_T'})^2)
  \nonumber \\
&= x \sum_a H_{aa}(Q^2,\mu^2)\int \df^2 {\bm k_T} \, 
  \omega \,f^{a}(x,{\bm k_T^2})\ D^{a}(\zh,(- \zh (\qT + \kT))^2)   \label{eq:def-convolution-integral-hadron-frame}
\,,\end{align}
The first two lines, \eq{def-convolution-integral}, describe the convolution in the photon–hadron frame, while the last two lines, \eq{def-convolution-integral-hadron-frame}, define the convolution in the hadron–hadron frame. Here, $\omega$ is a weight function, which in general depends on the transverse momenta of the incoming and outgoing quarks in the Trento frame, ${\bm k_\perp}$, ${\bm p_\perp}$, or in the hadron frame, ${\bm k_T}$, ${\bm p_T'}$, and $\bfhp = \bfPhperp/\Phperp=-\qT/q_T$. The sum runs over all quark and antiquark flavors $a=u,\bar u,d,\bar d$, etc. The hard function\footnote{Notice that generally, the hard function may be off-diagonal, $H_{aa'}$, in particular at N3LL. We will consider the lowest order in which it is diagonal and trivial, $H_{aa}(Q^2,Q^2)=e_a^2$.} for the SIDIS process is denoted by $H_{aa}$. Each TMD obeys TMD evolution equations and depends on two scales (not shown in the equations above), corresponding to the regulator for the ultraviolet divergence, $\mu$, and the regulator for the rapidity divergence, $\zeta$; see Ref.~\cite{Collins:2011zzd,Boussarie:2023izj}. In the following, the flavor index $a$ and the scale dependence of TMDs will be omitted in our formulas.

Notice that according to Refs.~\cite{Boer:2003cm,Meissner:2010cc,Collins:2011zzd,Metz:2016swz,Boussarie:2023izj} the density interpretation of TMD FFs requires that they depend on ${\bm p_\perp^2}$ in photon-hadron frame or on $(-\zh {\bm p_T'})^2$ in hadron-hadron frame, it is also evident in our \eqs{def-convolution-integral}{def-convolution-integral-hadron-frame}. In the following Section we will indicate this dependence explicitly.

\subsection{The hadronic tensor}
\label{sec:tensor}
Information on the structure of the target is encoded in the hadronic tensor.
Leading power and next-to-leading-power hadronic tensor for SIDIS was introduced  by Mulders and Tangerman~\cite{Mulders:1995dh} and Kotzinian~\cite{Kotzinian:1994dv}, and next-to-leading-power SIDIS hadroninc tensor was studied by Ebert, Gao, and Stewart in Ref.~\cite{Ebert:2021jhy} and by Rodini and Vladimirov in Ref.~\cite{Rodini:2023plb}. Next-to-next-to-leading-power kinematic corrections were studied by Vladimirov in Ref.~\cite{Vladimirov:2023aot} and by Pilo\~neta and Vladimirov in Ref.~\cite{Piloneta:2025jjb}.

Here we will outline how  to obtain power corrections to SIDIS hadronic tensor from the corresponding 
results for the Drell-Yan using the rapidity factorization formalism~\cite{Balitsky:2020jzt,Balitsky:2024ozy}.   In the Drell-Yan process two hadrons $A$ and $B$ collide and produce a detected lepton-antilepton pair 
\begin{align}
 p(P_A)+p(P_B) \to \ell^+(l) + \ell^-(l') +  X
\,.\end{align}
The center of mass energy is $s=(P_A+P_B)^2$. When the transverse component $q_T$ of the $\ell^+ \ell^-$ final state momentum, $q = l + l'$, is small compared to its invariant mass $Q^2 = q^2$, i.e. $q_T^2 \ll Q^2$, TMD factorization theorem is valid and the hadronic tensor contains TMD PDFs of hadrons $A$  and $B$. For Drell-Yan, it is natural to use the hadron directions as reference directions; therefore, one discusses the quark transverse momentum as relative to its parent hadron. 

We will refer to hadron $A$ as the projectile and to hadron $B$ as the target. 
In the methodology of Ref.~\cite{Balitsky:2017flc} quark and gluon fields of the target and projectile are separated into three sectors: ``projectile'' fields with small components along the target, ``target'' fields with small components
 along the projectile, and ``central'' fields which are neither projectile nor target ones.
 To obtain the TMD factorization, one 
 ``freezes'' the projectile and target fields and integrates over the central fields. The result of the integration is a series 
 of diagrams in the background of projectile and target fields. At the tree level, the sum of such diagrams describes a solution
 of QCD classical equations with sources being projectile and target fields.
 
The general solution, known as the scattering of two ``color glass condensates,'' is not known at present. However, if $q_T^2 / Q^2 \ll 1$, this ratio can be used as an expansion parameter to obtain the solution as a series of projectile and target TMD operators suppressed by powers of $Q^2$. At this stage, there is no difference between the Drell-Yan and SIDIS cases. The difference arises after integration over the projectile and target fields to yield the TMD matrix elements. The target matrix elements for SIDIS are the same as in Drell-Yan, except for the direction of the corresponding gauge links, which extend to $+\infty$ instead of $-\infty$ as in the Drell-Yan case. Regarding the projectile matrix elements, the TMD PDFs of the projectile are replaced by TMD fragmentation functions (TMD FFs). To remain consistent with the notations of the previous section, we will modify the conventions of Refs.~\cite{Balitsky:2020jzt,Balitsky:2024ozy} by setting $\perp \to T$ and $x \to b$.

We use Sudakov variables $\alpha$ and $\beta$ to parametrize
\begin{equation}
q = \alpha \, p_1 + \beta \, p_2 + q_T,
\end{equation}
where $p_1$ and $p_2$ are light-like vectors. In Drell-Yan,
\begin{equation}
P_A = p_1 + \frac{M_A^2}{s}\, p_2, \qquad P_B = p_2 + \frac{M_B^2}{s}\, p_1.
\end{equation}
We neglect hadron masses, i.e., $P_A^2 = P_B^2 = 0$, so that $p_1^\mu = P_A^\mu$ and $p_2^\mu = P_B^\mu$. In Semi-Inclusive Deep Inelastic Scattering (SIDIS), we take $p_1^\mu = P_h^\mu$ and $p_2^\mu = P^\mu$.

We use the following notation for two-dimensional transverse vector products:
\begin{equation}
(a,b)_T \equiv \mathbf{a}_T \cdot \mathbf{b}_T.
\end{equation}
We define
\begin{equation}
\varsigma \equiv 2 (p_1 \cdot p_2),
\end{equation}
so that in Drell-Yan $\varsigma = 2(P_A \cdot P_B) = s$, while in SIDIS $\varsigma = 2 (P \cdot P_h) = Q^2 \, z_h / x$.
For Drell-Yan, $\alpha \equiv x_1$ and $\beta \equiv x_2$, which leads to the standard relation
\begin{equation}
q^2 \equiv Q^2 = x_1 x_2 s - \mathbf{q}_T^2 > 0.
\end{equation}
In SIDIS,
\begin{equation}
\alpha = \frac{1}{z_h}, \qquad \beta = -x \left( 1 - \frac{\mathbf{q}_T^2}{Q^2} \right),
\end{equation}
and one has
\begin{equation}
q^2 \equiv -Q^2 < 0.
\end{equation}

The metric tensor can be written in terms of the longitudinal and transverse parts as follows
\begin{align}
g^{\mu\nu} &= g^{\mu\nu}_{\parallel}+g^{\mu\nu}_{T} = \frac{p_1^\mu p_2^\nu + p_2^\nu p_1^\mu}{(p_1\cdot p_2)}+g^{\mu\nu}_{T} \, ,\\
~{}^{\rm DY}g^{\mu\nu}_{\parallel} &= \frac{2 }{s}(p_1^\mu p_2^\nu + p_2^\nu p_1^\mu) \; , \\
~{}^{\rm SIDIS}g^{\mu\nu \rm}_{\parallel} &= \frac{2 x}{\zh Q^2}(p_1^\mu p_2^\nu + p_2^\nu p_1^\mu) \;.
\end{align}
We will use the transverse tensors $g^{\mu\nu}_{T}$ and $\epsilon^{\mu\nu}_{T}$, whose only nonzero components are $g^{11}_{T}=g^{22}_{T}=1$ and $\epsilon^{12}_{T}=-\epsilon^{21}_{T}=1$.

%
Quark TMD PDFs are defined by the following  correlator, see e.g. Ref.~\cite{Bacchetta:2006tn,Boussarie:2023izj}, 
\begin{eqnarray}
&&\hspace{-11mm}
\Phi_{ij}(x,{\bfkperp})~=~\int\! \frac{\df b^-\df^2 b_T}{8\pi^3}e^{-ixP^+b^- +i(k,b)_T}\langle P|\bsi_j(b)[b,\infty][\infty,0]\psi_i(0)|P\rangle
\nonumber\\
&&\hspace{-11mm}
=~\half\left[f_1(x,{\bfkperp})\sln_++ih_1^\perp{[\slk_T,\sln_+]\over 2M}\right]_{ij}+{M\over 2P^+}\left[e+f^\perp{\slk_T\over M}
-{g^\perp \gamma_5\over M}\epsilon_T^{\rho\sigma}\gamma_\rho k_{T\sigma}
+{i\over 2}h[\sln_+,\sln_-]\right]_{ij} \,
\label{eq:correlator}
\end{eqnarray}
where we  include only the functions related to the unpolarized scattering that we study in this paper and indicate the functional dependence only for the unpolarized TMD, $f_1$, for brevity.
In \eq{correlator}  two light-like vectors $n_{+}$ and $n_{-}$ are used, such that for any four vector $a$ one has $a^+ = a\cdot n^{-}, a^{-}
= a\cdot  n^{+}$ and $a_T \cdot n^+ = a_T\cdot n^- = 0$. They are related to our Sudakov vectors as follows:
\begin{align}
n_+^\mu \equiv \frac{p_2^\mu}{\sqrt{\varsigma/2}} \;\;, n_-^\mu \equiv \frac{p_1^\mu}{\sqrt{\varsigma/2}} \;.
\end{align}
Notice that in the frame we consider for SIDIS
\begin{align}
P^+ = P \cdot n_{-} = \sqrt{\varsigma/2} \;\;, P_h^- = P_h \cdot n_{+} = \sqrt{\varsigma/2} \,.
\end{align}
The leading twist unpolarized TMD is projected from the correlator as follows:
\begin{align}
f_1(x,{\bfkperp^2}) = \frac{1}{2}\;{\rm Tr}\biggl[\gamma^+ \;\Phi(x,{\bfkperp})\biggr]\; ,
\end{align}
while the subleading TMDs are obtained by projecting with $\mathbbm{1}$ and $\gamma^i$. 

The correlator for the fragmentation functions reads 
%
\begin{eqnarray}
&&\hspace{-14mm}
\Delta_{ij}(\zh,\bfkperp)={1\over 2 \zh N_c}\!\int\! {\df b^+\df^2 b_T\over 8\pi^3}e^{i P^-/\zh b^+ -i(k,b)_T}\sum_X\langle 0|\psi_i(b)|P_h+X\rangle\langle P_h+X|\bsi_j(0)|0\rangle
\nonumber\\
&&\hspace{-14mm}
=\half\Big[D_1\left(\zh,\zh^2 \kT^2\right)\sln_- +iH_1^\perp{[\slk_T,\sln_-]\over 2m_N}\Big]+{m_N\over 2P^-}\Big[E+F^\perp{\slk_T\over m_N}
+{G^\perp \gamma_5\over m_N}\epsilon_T^{\rho\sigma}\gamma_\rho k_{T\sigma}
+{i\over 2}H[\sln_-,\sln_+]\Big] \label{eq:ff_fourier}
\end{eqnarray}
where the functional dependence is indicated only for $D_1$ for brevity.
In SIDIS the produced hadron moves fast in the ``$-$" light-cone direction, and the twist-2 TMD FF is projected out as
\begin{align}
 D_1(\zh,\zh^2\bfkperp^2) =  \frac{1}{2}{\rm Tr}\big[\gamma^-\Delta(\zh,\bfkperp)\big]\; , 
\end{align}
while the subleading TMD FFs are obtained by projecting with $\mathbbm{1}$ and $\gamma^i$. Recall that, as derived in Ref.~\cite{Collins:2011zzd}, the dependence of the fragmentation functions on $\zh^2\bfkperp^2$ is consistent with the number density interpretation of the fragmentation functions. Accordingly, this dependence is standard in phenomenological applications~\cite{Boussarie:2023izj}.

For the Drell-Yan process, the large-$N_c$ power corrections up to $\mathcal{O}(1/Q^2)$ were obtained in Ref.~\cite{Balitsky:2024ozy} by using the QCD equations of motion to convert quark–antiquark–gluon ($\bar q F q$) TMDs into quark–quark ($\bar q q$) TMDs. For example, the QCD Dirac equation of motion (EOM)
\begin{eqnarray}
&&\hspace{-1mm}
\bsi(b)\slA_T(b)~=~i\partial_i\bsi(b)\gamma^i+i\sqrt{2\over \varsigma}\partial_+\bsi(b)\slp_1
+i\sqrt{2\over \varsigma} \bsi(b)  \stackrel{\leftarrow}D_+\slp_2
\label{eq:mp}
\end{eqnarray}
leads to 
\begin{eqnarray}
&&\hspace{-1mm}
{\sqrt{2/ \varsigma}\over 16\pi^3}\!\int\! \df b^+ \df^2\bT~e^{-i\alpha\sqrt{\varsigma/2} b^+ +i(q-k,b)_T}
\langle p_A|{\bar\psi}(b^+,\bT)\slA(b^+,\bT)\slp_B\gamma_i\psi(0)|p_A\rangle
\nonumber\\
&&\hspace{-1mm}
=~(q-k)_i\big\{f_1(\alpha,(\qT-\kT)^2)-\alpha \big[ f^\perp(\alpha,(\qT-\kT)^2)+ig^\perp(\alpha,(\qT-\kT)^2)\big]\big\}
\label{eq:ft}
\end{eqnarray}
and
\begin{eqnarray}
&&\hspace{-1mm}
{\sqrt{2/ \varsigma}\over 16\pi^3}\!\int\! \df b^+ \df^2\bT~e^{-i\alpha\sqrt{\varsigma/2} b^+ +i(q-k,b)_T}
\langle p_A|{\bar\psi}(b^+,\bT)\slA(b^+,\bT)\slp_B\psi(0)|p_A\rangle
\nonumber\\
&&\hspace{0mm}
=~\big\{-i{(\qT-\kT)^2\over M} h_1^\perp(\alpha,(\qT-\kT)^2)-M\alpha \big[ e(\alpha,(\qT-\kT)^2)+ih(\alpha,(\qT-\kT)^2)\big]\big\}
\label{eq:fth}
\end{eqnarray}
The real part in brackets on the right-hand side of \eq{ft}, $f_1 - \alpha f^\perp$, is usually denoted as $-\alpha \tilde f^\perp$, while the imaginary part in \eq{fth}, $-(\qT-\kT)^2/M, h_1^\perp - M\alpha h$, is denoted as $-\alpha M \tilde h$.
These tilde terms are neglected in the Wandzura--Wilczek (WW) approximation. Terms of this type are important for the restoration of electromagnetic (EM) gauge invariance, as discussed in Ref.~\cite{Balitsky:2020jzt}. A nontrivial observation made in Ref.~\cite{Balitsky:2020jzt} for the Drell–Yan process is that, when all contributions originating from $\bar{\psi}(x)\slashed{A}(x)$ and $\slashed{A}(0)\psi(0)$ are properly combined, one obtains an EM gauge-invariant contribution involving the functions $f_1$ and $h_1^{\perp}$. We refer to this result as the gauge completion of the leading-twist contribution.

In contrast, for the terms proportional to $f^{\perp}$ and $h$, achieving gauge invariance requires corrections of order $1/Q^3$, as discussed in Ref.~\cite{Balitsky:2020jzt}. It is therefore natural to examine the assumption that the gauge completion of leading-twist functions is numerically more important than that of higher-twist functions. A comparison with experimental data on $Z$-boson angular coefficients~\cite{Balitsky:2021fer} indicates that this approximation is indeed reasonable.

The equations of motion of the type given in Eqs.~(\ref{eq:mp}, \ref{eq:fth}) are the same for the SIDIS process. However, the Fourier transform of the fragmentation functions defined in \eq{ff_fourier} is somewhat different; therefore, instead of Eq.~(\ref{eq:ft}), one obtains the following relations.
\begin{eqnarray}
&&\hspace{-1mm}
{\sqrt{2/ \varsigma}\over 16\pi^3}{\alpha\over 2N_c}\!\int\! \df b^+ \df^2\bT~e^{i\alpha\sqrt{\varsigma/2} b^+ -i(q+k,b)_T}
\sum_X\langle 0|{\bar\psi}(b^+,\bT)\slA(b^+,\bT)|p_A+X\rangle\langle p_A+X|\slp_B\gamma_i\psi(0)|N\rangle
\nonumber\\
&&
\nonumber\\
&&\hspace{2mm}
=~-(q+k)_i\left(\barD_1\left({1\over\alpha},{1\over \alpha^2}(\qT+\kT)^2\right)-\alpha\left[ \barF^\perp-i\barG^\perp\right]\right)
\label{eq:ftD}
\end{eqnarray}
and
\begin{eqnarray}
&&\hspace{-1mm}
{\sqrt{2/ \varsigma}\over 16\pi^3}{\alpha\over 2N_c}\!\int\! \df b^+ \df^2 \bT~e^{i\alpha\sqrt{s/2} b^+ -i(q+k,b)_T}
\sum_X\langle 0|{\bar\psi}(b^+,\bT)\slA(b^+,\bT)|p_A+X\rangle\langle p_A+X|\slp_B\psi(0)|N\rangle
\nonumber\\
&&
\nonumber\\
&&\hspace{-2mm}
=~~\bigg[i{(\qT+\kT)^2\over \mh}\barH_1^\perp\Big({1\over\alpha},{1\over \alpha^2}(\qT+\kT)^2\Big)
+\alpha \mh\left(\barE+i\barH\right)\bigg]
\label{eq:ftH}
\end{eqnarray}
Here for brevity we indicate the functional dependence only for $D_1$ and $H_1^\perp$.

If one considers the analytic continuation of ${}^{\rm DY}W_{\mu\nu}(-q)$ : 
\begin{eqnarray}
&&\hspace{-1mm}
{\sqrt{2/\varsigma}\over 16\pi^3}\!\int\! \df b^- \df^2 \bT~e^{i\alpha\sqrt{\varsigma/2} b^- -i(q+k,b)_T}
\langle p_A|{\bar\psi}(b^-,\bT)\slA(b^-,\bT)\slp_B\gamma_i\psi(0)|p_A\rangle
\label{eq:ftDfromDY}\\
&&
\nonumber\\
&&\hspace{1mm}
=~-(q+k)_i\left(f_1(-\alpha,(-\qT-\kT)^2)+\alpha \left[ f^\perp(-\alpha,(-\qT-\kT)^2)+ig^\perp(-\alpha,(-\qT-\kT)^2)\right]\right),
\nonumber
\end{eqnarray}
the Fourier transform has the same structure as the one in Eq.~(\ref{eq:ftD}).  
Similarly, the analytic continuation of \eq{fth} gives
\bega
&&\hspace{-1mm}
{\sqrt{2/ \varsigma}\over 16\pi^3}\!\int\! \df b^+ \df^2 \bT~e^{i\alpha\sqrt{s/2} x^+ -i(q+k,x)_\perp}
\langle p_A|{\bar\psi}(b^+,\bT)\slA(b^+,\bT)\slp_B\psi(0)|p_A\rangle
\\
&&\hspace{-3mm}
=~\left[-i{(\qT+\kT)^2\over M}h_1^\perp(-\alpha,(-\qT-\kT)^2)+\alpha M \big[ e(-\alpha,(-\qT-\kT)^2)+ih(-\alpha,(-\qT-\kT)^2)\big]\right] \nonumber
\label{9.22cont}
\end{eqnarray}
which has the same structure as \eq{ftH}.

In general, it can be demonstrated that the same formulas used for the Drell–Yan case can also be applied to SIDIS via the following procedure: one constructs the SIDIS hadronic tensor from Ref.~\cite{Balitsky:2024ozy} by analytically continuing the Drell–Yan hadronic tensor to negative momentum, $q \to -q$, replacing the hadron mass $M \to m_h$, and substituting the TMD PDFs of the projectile with TMD fragmentation functions (TMD FFs) according to the following rules:
\begin{align}
 f_1(-\alpha,(-\kT)^2)\rightarrow \barD_1\left({1\over\alpha},\frac{\kT^2}{\alpha^2}\right){2N_c\over\alpha},~f^\perp(-\alpha,(-\kT)^2)\rightarrow -\barF^\perp\left({1\over\alpha},\frac{\kT^2}{\alpha^2}\right){2N_c\over\alpha},\nonumber \\
 ~g^\perp(-\alpha,(-\kT)^2)\rightarrow -\barG^\perp\left({1\over\alpha},\frac{\kT^2}{\alpha^2}\right){2N_c\over\alpha}, \\
 \barf_1(-\alpha,(-\kT)^2)\rightarrow D_1\left({1\over\alpha},\frac{\kT^2}{\alpha^2}\right){2N_c\over\alpha},~\barf^\perp(-\alpha,(-\kT)^2)\rightarrow -F^\perp\left({1\over\alpha},\frac{\kT^2}{\alpha^2}\right){2N_c\over\alpha},\nonumber \\
 ~\bar g^\perp(-\alpha,(-\kT)^2)\rightarrow -G^\perp\left({1\over\alpha},\frac{\kT^2}{\alpha^2}\right){2N_c\over\alpha}.
 \end{align}

and
\begin{align}
\barh_1^\perp(-\alpha,(-\kT)^2)\rightarrow -H_1^\perp\Big({1\over\alpha},\frac{\kT^2}{\alpha^2}\Big){2N_c\over\alpha},~~~\barh(-\alpha,(-\kT)^2)\rightarrow H^\perp\Big({1\over\alpha},\frac{\kT^2}{\alpha^2}\Big){2N_c\over\alpha}
,\nonumber \\
~~~\bar e(-\alpha,(-\kT)^2)_\perp\rightarrow -E\left({1\over\alpha},\frac{\kT^2}{\alpha^2}\right){2N_c\over\alpha} \;, \\
h_1^\perp(-\alpha,(-\kT)^2)\rightarrow -\barH_1^\perp\Big({1\over\alpha},\frac{\kT^2}{\alpha^2}\Big){2N_c\over\alpha},~~~\barh(-\alpha,(-\kT)^2)\rightarrow H^\perp\Big({1\over\alpha},\frac{\kT^2}{\alpha^2}\Big){2N_c\over\alpha}
,\nonumber \\
~~~e(-\alpha,(-\kT)^2)_\perp\rightarrow -\barE\left({1\over\alpha},\frac{\kT^2}{\alpha^2}\right){2N_c\over\alpha} \;.
\end{align}

The complete result for $\mathcal{O}(1/ Q^2)$ and leading-$N_c$ power corrections for the DY hadronic tensor reads~\cite{Balitsky:2024ozy}
\beq
{}^{\rm DY}W_{\mu\nu}(q)~=~\sum_a e_a^2~\big[W^{1}_{\mu\nu}(q)+W^{2}_{\mu\nu}(q)
+W^{3}_{\mu\nu}(q)\big]~+~\mathcal{O}\left({1\over Q^3}\right)~+~\mathcal{O}\left({1\over N_c^2}\right)
\label{dyfinalresult}
\eeq
The first, EM-gauge invariant part $W^{1}$ is a ``gauge completion'' of the leading-twist result:
\begin{eqnarray}
&&\hspace{-1mm}
{}^{\rm DY}W^1_{\mu\nu}(q)~=\frac{1}{N_c}\!\int\!\df^2\kT \Big({}^{\rm DY}\calw^F_{\mu\nu}(q,\kT)
[f_1(\alpha,\qT-\kT)\barf_1(\beta,\kT)+f_1\leftrightarrow\barf_1]
\nonumber\\
&&\hspace{-1mm}
+~{}^{\rm DY}\calW^H_{\mu\nu}(q,\kT)\{h_1^\perp(\alpha,\qT-\kT)\barh_1^\perp(\beta,\kT)
+h_1^\perp\leftrightarrow\barh_1^\perp\}
\Big)
\label{dyresult1}
\end{eqnarray}
where the transverse structures ${}^{\rm DY}\calW^{F}_{\mu\nu}(q,\kT)$ and ${}^{\rm DY}\calW^{H}_{\mu\nu}(q,\kT)$ are given by Eqs. (7.3) and (7.4) of Ref.~\cite{Balitsky:2024ozy}, and the flavor index $a$ is implicit for all TMD functions. This term was obtained by applying the equations of motion and contains only products of leading-twist functions, such as $f_1$ and $h_1^\perp$.

The second term $W^{2}$ contains mixed products of leading-twist functions,  $f_1$ and $h_1^\perp$, and quark-antiquark TMDs of subleading twist, such as $f^\perp$, $h$, etc. This term is also EM gauge invariant.

The last contribution $W^{3}$, which turns out to be the only one of the three that is not gauge invariant, appears in $W^{3}_{\mu\nu}(q)$ and is composed of quark-quark-gluon TMDs that cannot be reduced to quark-antiquark TMDs by EOMs. 

The number of different TMDs appearing in the three terms is about 30, so it is useful to understand which ones are the most important numerically. To this end, the assumption used in Refs.~\cite{Balitsky:2017gis,Balitsky:2021fer} consists of systematically neglecting products of TMDs that mix leading and subleading twist as well as neglecting those quark-quark-gluon TMDs that cannot be related to leading twist TMDs via equations of motion, while  retaining only products of leading-twist TMDs, $f_1 \bar f_1$ and $h_1^\perp \bar h_1^\perp$. This means that we assume that the most significant contribution to the hadronic tensor in Eq.~(\ref{dyfinalresult}) is $W^{1}$ and the other two terms, $W^{2}$ and $W^{3}$, can be neglected.

It was shown in Refs.~\cite{Balitsky:2017gis,Balitsky:2021fer} that this assumption allows for a good description of the angular distributions of $Z$-boson DY production by numerical comparison with LHC DY data: the DY coefficients, see Refs.~\cite{Balitsky:2021fer}, $A_0$ and $A_2$ agree at the level of ${1/N_c}\sim 30\%$ accuracy, while other angular coefficients, which are zero under naive estimates, are experimentally found to be an order of magnitude smaller than $A_0$ and $A_2$.

In this paper, we aim to test a similar assumption. We take into account the leading-twist distributions $f_1, D_1$ and $h_1^\perp, H_1^\perp$ along with their gauge completions. That is, in Eq.~(\ref{eq:ftDfromDY}) only $f_1$ is retained. Under these assumptions, the SIDIS hadronic tensor takes the form
\beq
\hspace{-0mm}
{}^{\rm SIDIS}W^1_{\mu\nu}(q)~=~2 \zh\!\int\!\df^2\kT \left(~{}^{\rm SIDIS}\calW^F_{\mu\nu}(q,\kT)\{D_1f_1+\barD_1 \barf_1\}
-{}^{\rm SIDIS}\calW^H_{\mu\nu}(q,\kT)\kappa\{H_1^\perp h_1^\perp+\barH^\perp \barh_1^\perp\}
\right)
\label{sidresult1}
\eeq
where $\kappa = -1$ for SIDIS and $\kappa = 1$ for Drell-Yan, see Ref.~\cite{Boussarie:2023izj}, such that $\kappa h_1^\perp$ always corresponds to Boer-Mulders function in a particular process:
\begin{align}
    h_1^{\perp \rm (DY)} = h_1^\perp\; \text{and}\;
     h_1^{\perp \rm (SIDIS)} = -h_1^\perp\, .
\end{align}
As introduced in TMD handbook~\cite{Boussarie:2023izj}, the notation for Boer-Mulders functions with explicit $\kappa$ allows one to avoid potential confusion and mismatches in formulas pertaining to the DY and SIDIS cases. Therefore, we will retain $\kappa h_1^\perp$ in the text for Boer-Mulders functions.

For a generic TMD PDF ${\mathcal F}$ and TMD FF ${\mathcal D}$ one has
\bega
&&\hspace{-1mm}
{\mathcal D} {\mathcal F}+\bar  {\mathcal D}\bar {\mathcal F}~\equiv~{\mathcal D}\left(\frac{1}{\alpha},{\frac{1}{\alpha^2}}( \qT+\kT)^2\right) {\mathcal F}(-\beta,\kT^2)
+\bar {\mathcal D}\left(\frac{1}{\alpha},{\frac{1}{\alpha^2}}(\qT+\kT)^2\right)\bar {\mathcal F}(-\beta,\kT^2)
\nonumber\\
&&\hspace{-1mm}
=~{\mathcal D}(\zh,{\zh^2}(\qT+\kT)^2){\mathcal F}(x,\kT^2)
+\bar{\mathcal D}(\zh,{\zh^2}(\qT+\kT)^2)\bar {\mathcal F}(x,\kT^2)\; .
\label{sidnotation}
\ega
The kinematical tensor structures $\calW^{F,H}_{\mu\nu}$ are taken from  Ref. \cite{Balitsky:2024ozy} and tensors for SIDIS are related to those in DY as follows 
\begin{eqnarray}
&&\hspace{-11mm}
~{}^{\rm SIDIS}\calW^{F}_{\mu\nu}(q,\kT)~=~{}^{\rm DY}\calW^{F}_{\mu\nu}(-q,\kT)~=~
-g_{\mu\nu}^T -{1\over Q^2}(q^\parallel_\mu q_{T\nu}+q^\parallel_\nu q_{T\mu})
+{{\bm q_T^2}\over Q^4}q^\parallel_\mu q^\parallel_\nu
\nonumber\\
&&\hspace{2mm}
+~{\tilq_\mu\tilq_\nu\over Q^4}[{\bm q_T^2}+4(k,q+k)_T]
+\left[{\tilq_\mu\over  Q^2}\left(g^T_{\nu i}+{q^\parallel_\nu q_i\over Q^2}\right)(q+2k)_T^i
+\mu\leftrightarrow\nu\right] 
\label{eq:wfsid}
\end{eqnarray}
%
\begin{eqnarray}
&&\hspace{-1mm}
~{}^{\rm SIDIS}\calW_{\mu\nu}^H(q,\kT)~=~{}^{\rm DY}\calW_{\mu\nu}^H(-q,\kT)~
\label{eq:whsid}\\
&&\hspace{-1mm}
=~{1\over {M m_h}}\left[k_{T\mu}(q+k)_{T\nu}+k_{T\nu}(q+k)_{T\mu}+g_{\mu\nu}^T(k,q+k)_T\right]
+2{\tilq_\mu\tilq_\nu-q^\parallel_\mu q^\parallel_\nu \over {Q^4Mm_h}}\kT^2(q+k)_T^2
\nonumber\\
&&\hspace{-1mm}
-~{1\over {Q^2Mm_h}}\left(q^\parallel_\mu\left[-\kT^2(q+k)_{T\nu}+k_{T\nu}(q+k)_T^2\right]
+~\tilq_\mu\left[-\kT^2(q+k)_{T\nu}-k_{T\nu}(q+k)_T^2\right]+\mu\leftrightarrow\nu\right)
\nonumber\\
&&\hspace{-1mm}
+~{\tilq_\mu\tilq_\nu+q^\parallel_\mu q^\parallel_\nu \over {Q^4Mm_h}}\left[{\bm q_T^2}+2(k,q+k)_T\right](k,q+k)_T
-~{q^\parallel_\mu\tilq_\nu+ \tilq_\mu q^\parallel_\nu \over {Q^4M m_h}}(2k+q,q)_T(k,q+k)_T 
\nonumber
\end{eqnarray}
Here $q^\parallel_\mu\equiv p_1/\zh-x p_2$ and $\tilq_\mu\equiv p_1/\zh+x p_2$.  
\footnote{In SIDIS, in our approximation $p_1^\mu = P_h^\mu$ and $p_2^\mu = P^\mu$. Strictly speaking,  $Q^2=-q_\parallel^2+{\bm q_T^2}$ and $-\beta=x+x{{\bm q_T^2}\over Q^2}$ but 
taking into account terms $\sim {\bm q_T^2}$ will lead to power corrections $\sim {1/Q^3}$.}
Notice that the leading-twist
contribution is given by the first terms in the R.H.S. of Eq.~(\ref{eq:wfsid}) and Eq.~(\ref{eq:whsid}).

It is easy to see that $~{}^{\rm SIDIS}W^1_{\mu\nu}$ is transverse 
\footnote{As discussed in Ref.~\cite{Balitsky:2024ozy}, the second part  $W^2_{\mu\nu}$  is also gauge invariant, but $W^3_{\mu\nu}$ is not. To achieve  the ``gauge completion'' for  $W^3_{\mu\nu}$ one needs to take into account ${1/Q^3}$ corrections.}
\begin{align}
q^\mu ~{}^{\rm SIDIS}W^1_{\mu\nu}(q) = q^\nu ~{}^{\rm SIDIS}W^1_{\mu\nu}(q) = 0 ,
\end{align}
As we mentioned above, our numerical estimates for the structure functions will be based solely on this term, so that $~{}^{\rm SIDIS}W_{\mu\nu} = ~\sum_a e_a^2 ~{}^{\rm SIDIS}W^1_{\mu\nu}$.

\section{Structure functions}
\label{sec:structure}
To obtain the structure functions that appear in \eq{SIDIS-leading} we use Eqs.~(2.85-2.48) from Ref.~\cite{Piloneta:2025jjb}, 
\begin{align} 
 F_{UU,T}	&= \frac{x}{4\zh}\left( {\mathcal S}^{\mu\nu}_1 -{\mathcal S}^{\mu\nu}_0\right)W_{\mu\nu} = \frac{1}{2} F_{UU,L} - \frac{x}{4\zh}{\mathcal S}^{\mu\nu}_0 W_{\mu\nu}\, ,\\
F_{UU,L}	&= \frac{x}{4\zh} 2{\mathcal S}^{\mu\nu}_1 W_{\mu\nu}\, ,\\
F_{UU}^{\cos\phi_h}
	&=
	\frac{x}{4\zh}{\mathcal S}^{\mu\nu}_3W_{\mu\nu}\, ,\\
 F_{UU}^{\cos 2\phi_h} &= \frac{x}{4\zh}\left( {\mathcal S}^{\mu\nu}_0 -{\mathcal S}^{\mu\nu}_1 - 2 {\mathcal S}^{\mu\nu}_2\right)W_{\mu\nu} = -\frac{1}{2} F_{UU,L} + \frac{x}{4\zh}\left( {\mathcal S}^{\mu\nu}_0  - 2 {\mathcal S}^{\mu\nu}_2\right)W_{\mu\nu}\, .
 \end{align}
Here, in our notations, we have
 \begin{align} 
 {\mathcal S}^{\mu\nu}_0&=g^{\mu\nu} - \frac{q^\mu q^\nu}{q^2}\; ,\\
 {\mathcal S}^{\mu\nu}_1&=\frac{(\tilq^\mu +\breq^\nu)(\tilq^\nu +\breq^\mu)}{-q^2}\; ,\\
{\mathcal S}^{\mu\nu}_2&=\frac{\breq^\mu \breq^\nu}{-{\bm q_T^2}}\; ,\\
{\mathcal S}^{\mu\nu}_3&=\frac{(\tilq^\mu +\breq^\mu)\breq^\nu + \mu\leftrightarrow \nu}{q_T Q}\; , 
 \end{align}
where the vectors are defined as 
\begin{align}
\tilq^\mu &= \alpha p_1^\mu - \beta p_2^\mu = \frac{P_h^\mu}{\zh} +x\left (1-\frac{\bm q_T^2}{Q^2}\right)P^\mu \; ,\\
\breq^\mu &= q_T^\mu +\frac{2 {\bm q_T^2}}{Q^2} p_2^\mu = q_T^\mu + \frac{2\bm q_T^2}{Q^2} P^\mu\, .
\end{align}
Using our ``gauge completion of the leading twist'' approximation, we obtain
\begin{align} 
 F_{UU,T}	&= x \sum_a  H_{aa}(Q^2,\mu^2)\int d^2\bfkperp \biggl[ \left( 1- \frac{2 \qT\cdot \bfkperp}{Q^2}\right)f_1(x,\bfkperp^2) D_1(\zh,((\qT+\bfkperp)\zh)^2) +
 \nonumber \\
 -&\frac{2 \bfkperp^2}{M m_h Q^2} \qT \cdot (\qT + \bfkperp)  h_1^\perp (x,\bfkperp^2)H_1^\perp(\zh,((\qT+\bfkperp)\zh)^2)\biggr] 
\,,\label{eq:FUU_Ian}\\
F_{UU,L}	&= x \sum_a  H_{aa}(Q^2,\mu^2)\int d^2\bfkperp \biggl[  \frac{4 \bfkperp^2}{Q^2} f_1(x,\bfkperp^2) D_1(\zh,((\qT+\bfkperp)\zh)^2) \nonumber \\
+& \frac{4 \bfkperp^2}{M m_h Q^2} \bfkperp \cdot (\qT + \bfkperp) h_1^\perp (x,\bfkperp^2)H_1^\perp(\zh,((\qT+\bfkperp)\zh)^2) \biggr]
\,,\label{eq:FUUL_Ian}\\
F_{UU}^{\cos\phi_h}
	&=
	x \sum_a  H_{aa}(Q^2,\mu^2)\int d^2\bfkperp \biggl[ \frac{2 \qT \cdot\bfkperp^{ }}{Q q_T} f_1(x,\bfkperp^2) D_1(\zh,((\qT+\bfkperp)\zh)^2)
    \nonumber \\
    +& \frac{2\bfkperp^2}{M m_h Q q_T}
	 \qT \cdot (\qT + \bfkperp) h_1^\perp (x,\bfkperp^2)H_1^\perp(\zh,((\qT+\bfkperp)\zh)^2) \biggr] ,
	\label{eq:FUUcosphi_Ian}\\
 F_{UU}^{\cos 2\phi_h} &= x \sum_a H_{aa}(Q^2,\mu^2)\int d^2\bfkperp \biggl[- \frac{2\qT\cdot \bfkperp}{Q^2}f_1(x,\bfkperp^2) D_1(\zh,((\qT+\bfkperp)\zh)^2) 
 \nonumber \\
 & + \biggl(
-\frac{\bfkperp \cdot (\qT + \bfkperp)}{M m_h} 
+\frac{2\bfkperp\cdot\qT \; ( \qT \cdot (\qT + \bfkperp))}{{\bf q}_T^2 M m_h} 
\nonumber \\
 &
 -\frac{2\bfkperp^2\; ( \qT \cdot (\qT + \bfkperp))}{Q^2 M m_h} 
	 \biggr)  h_1^\perp (x,\bfkperp^2)H_1^\perp(\zh,((\qT+\bfkperp)\zh)^2) \biggr]
\,,\label{eq:FUUcos2phi_Ian}\end{align}

Notice that in the formulas above, Boer-Mulders functions are for Drell-Yan.

We now rewrite Eqs.~(\ref{eq:FUU_Ian}-\ref{eq:FUUcos2phi_Ian}) in the standard notations of Eqs.~(\ref{eq:def-convolution-integral}) using Eq.~(\ref{eq:approx}), $-{\bm p_\perp}/\zh = \qT+\bfkperp$ and $\bfhp = -\qT/q_T$, and using Boer-Mulders functions for SIDIS, $\kappa h_1^\perp$ with $\kappa=-1$,
\begin{align} 
 F_{UU,T}	&= {\cal C} \biggl[ \left( 1+\frac{2 q_T }{Q^2} (\bfhp\cdot {\bm k_\perp})\right) f_1(x,{\bm k_\perp^2}) D_1(\zh,{\bm p_\perp^2})  
 \nonumber \\
 &  +\frac{2 q_T {\bm k_\perp^2}}{\zh M m_h Q^2} (\bfhp \cdot {\bm p_\perp})  \kappa h_1^\perp(x,{\bm k_\perp^2}) H_1^\perp(\zh,{\bm p_\perp^2})\biggr] 
\,,\label{eq:FUU_Ian1}\\
F_{UU,L}	&= {\cal C} \biggl[  \frac{4 {\bm k_\perp^2}}{Q^2} f_1 D_1 + \frac{4 {\bm k_\perp^2}}{\zh M m_h Q^2} ({\bm k_\perp} \cdot {\bm p_\perp}) \kappa h_1^\perp H_1^\perp \biggr]
\,,\label{eq:FUUL_Ian1}\\
F_{UU}^{\cos\phi_h}
	&=
	{\cal C}\biggl[ -\frac{2 (\bfhp \cdot{\bm k_\perp})}{Q} f_1 D_1 - \frac{2{\bm k_\perp^2}}{\zh M m_h Q} ({\bfhp}\cdot{\bm p_\perp})
	\kappa h_1^\perp H_{1}^{\perp }
	\biggr] ,
	\label{eq:FUUcosphi_Ian1}\\
 F_{UU}^{\cos 2\phi_h} &= {\cal C}\biggl[ \frac{2q_T (\bfhp \cdot {\bm k_\perp})}{Q^2}f_1 D_1 
 \nonumber \\
 &  - \left(
\frac{({\bm k_\perp}\cdot{\bm p_\perp})}{\zh M m_h} 
-\frac{2(\bfhp\cdot{\bm k_\perp}) \; (\bfhp\cdot{\bm p_\perp})}{\zh M m_h} \right) \kappa h_1^\perp H_{1}^{\perp}
 +\frac{2 q_T {\bm k_\perp^2}\; }{\zh M m_h Q^2} (\bfhp\cdot{\bm p_\perp})
	  \kappa h_1^\perp H_{1}^{\perp}
	\biggr] 
\,,\label{eq:FUUcos2phi_Ian1}\end{align}
For the NNLP contributions, $\sim 1/Q^2$, in these equations one obtains the following relations:
\begin{align}
F_{UU,T}^{\rm NNLP} = F_{UU}^{\cos 2\phi_h \rm NNLP} = -\frac{q_T}{Q} F_{UU}^{\cos \phi_h}\; . \label{eq:relation}
\end{align}
TMDs are often studied in $b_T$ space and the convolutions can be written in terms of Fourier transformations. We express the convolutions in Eq.~\eqref{eq:FUU_Ian1},\eqref{eq:FUUL_Ian1},\eqref{eq:FUUcosphi_Ian1}, and \eqref{eq:FUUcos2phi_Ian1} through Fourier transforms of products of TMDs in $b_T$ space~\cite{Boer:2011xd},
where the Fourier transform is defined as, see Ref.~\cite{Boussarie:2023izj},
\begin{align} \label{eq:convBessSIDIS1}
 {\cal B}[\tilde f_{}^{(m)}\; \tilde D_{}^{(n)}] 
 \equiv  & x \sum_a H_{aa}(Q^2,\mu^2) \int_0^\infty \frac{\df b_T}{2\pi}\; b_T\, b_T^{m+n} \, J_{m+n}(q_T b_T) \nonumber \\
& \times \tilde f^{(m)}(x,{b_{T}}) \; 
 \tilde D^{(n)}(\zh,{b_{T}})
\,.\end{align}
The Fourier-transformed TMD PDFs $\tilde f$ and TMD FFs $\tilde D$ and their derivatives $\tilde f^{(n)}$ and  $\tilde D^{(n)}$ are defined as, see e.g.   Ref.~\cite{Boer:2011xd,Boussarie:2023izj},
\begin{align} \label{eq:TMD_bt_derivative}
 \tilde f^{(n)}(x, b_T) &
 \equiv n! \left(\frac{-1}{M^2 b_T} \partial_{b_T} \right)^n \tilde f(x, b_T)
\nonumber\\&
 = \frac{2\pi\, n!}{(M^2)^n} \int_0^\infty \df k_\perp \, k_\perp \left(\frac{k_\perp}{b_T}\right)^n J_n(b_T k_\perp) \, f(x, k_\perp)
\,, \\
\label{eq:TMDFF_bt_derivative}
 \tilde D^{(n)}(\zh , b_T) &
 \equiv n! \left(\frac{-1}{m_h^2 b_T} \partial_{b_T} \right)^n \tilde D(z, b_T)
\nonumber\\&
 = \frac{2\pi\, n!}{(m_h^2)^n} \int_0^\infty \frac{\df p_\perp \, p_\perp}{\zh^2} \left(\frac{p_\perp}{\zh b_T}\right)^n J_n\left(\frac{b_T p_\perp}{\zh}\right) \, D(\zh, p_\perp)
\,.
\end{align}
We obtain:

\begin{align}
 F_{UU,T}	&= {\cal B} [\tilde f_1^{(0)} \tilde D_1^{(0)}] +   \frac{2 q_T M^2}{Q^2}{\cal B} \left[\tilde f_1^{(1)} \tilde D_1^{(0)} \right] + 
\frac{2q_T \mh}{Q^2 M} {\cal B}\left[\widetilde{k_\perp^2 \kappa h_{1}^\perp}^{(0)} \tilde H_{1}^{\perp(1)}\right]\, 
\,,\label{eq:FUU_bspace_Ian} \\
{F}_{UU,L} &= \frac{4}{Q^2} {\cal B}\left[\widetilde{k_\perp^2 f_{1}}^{(0)}\, \tilde D_{1}^{(0)}\right] - 
\frac{4 \mh M}{Q^2} {\cal B'}\left[\widetilde{k_\perp^2 \kappa h_{1}^\perp}^{(1)}\, \tilde H_{1}^{\perp(1)}\right]\,
 \,,
\label{eq:FUUL-bspace_Ian} \\
 F_{UU}^{\cos \phi_h}	&= -\frac{2 M^2}{Q}{\cal B} \left[\tilde f_1^{(1)} \tilde D_1^{(0)} \right]- 
\frac{2 \mh}{Q M} {\cal B}\left[\widetilde{k_\perp^2 \kappa h_{1}^\perp}^{(0)}\, \tilde H_{1}^{\perp(1)}\right]\,,\label{eq:FUU_cosphi_bspace_Ian} \\
 F_{UU}^{\cos 2 \phi_h}	&= M\mh\, {\cal B}\left[\widetilde{\kappa h_{1}^{\perp}}^{(1)}\, \tilde H_{1}^{\perp (1)}\right]
+ \frac{2 q_T M^2}{Q^2}{\cal B} \left[\tilde f_1^{(1)} \tilde D_1^{(0)} \right] + 
\frac{2q_T \mh}{Q^2 M} {\cal B}\left[\widetilde{k_\perp^2 \kappa h_{1}^\perp}^{(0)}\, \tilde H_{1}^{\perp(1)}\right]
\,,\label{eq:FUU_cos2phi_bspace_Ian}
\end{align}
where we define a new convolution:
\begin{align} \label{eq:convBessSIDIS1new}
 {\cal B'}[\tilde f_{}^{(m)}\; \tilde D_{}^{(n)}] 
 \equiv  & x \sum_a H_{aa}(Q^2,\mu^2) \int_0^\infty \frac{\df b_T}{2\pi}\; b_T\, b_T^{m+n} \, J_{0}(q_T b_T) \nonumber \\
& \times \tilde f^{(m)}(x,{b_{T}}) \; 
 \tilde D^{(n)}(\zh,{b_{T}})
\,.\end{align}
The functions $\widetilde{k_\perp^2 f}^{(0)}$ and $\widetilde{k_\perp^2 f}^{(1)}$ are defined as
\begin{align}
\widetilde{k_\perp^2 f}^{(n)}(x, b_T) \equiv \frac{2\pi\, n!}{(M^2)^n}  \int_0^\infty \df k_\perp \, k_\perp \left(\frac{k_\perp}{b_T}\right)^n  J_n(b_T k_\perp) \, \; {\bm k_\perp^2} f(x, {\bm k_\perp^2}) \; .
\end{align}

Note that in this section we presented our results in a few different ways, Eqs.~(\ref{eq:FUU_Ian} - \ref{eq:FUUcos2phi_Ian}) are in the original formalism from Refs.~\cite{Balitsky:2017gis,Balitsky:2021fer}, while  Eqs.~(\ref{eq:FUU_Ian1} - \ref{eq:FUUcos2phi_Ian1}) are written in the standard notations of Ref.~\cite{Boussarie:2023izj}. Finally, it is convenient to express convolutions in the Fourier conjugate space, see Ref.~\cite{Boussarie:2023izj}, and  Eqs.~(\ref{eq:FUU_bspace_Ian} - \ref{eq:FUU_cos2phi_bspace_Ian}) are written in terms of convolutions in $b_T$ space. These formulas can facilitate numerical studies if TMDs are expressed in $b_T$ rather than $k_\perp$ space.

\section{Numerical estimates and comparisons}
\label{sec:numerics}
In this section, we provide numerical estimates for structure functions and compare our results with the existing literature, in particular with two recent papers: Piloñeta and Vladimirov~\cite{Piloneta:2025jjb} and Ebert, Gao, and Stewart~\cite{Ebert:2021jhy}, as well as earlier works, including Bacchetta et al.~\cite{Bacchetta:2006tn}, the generalized helicity formalism by Anselmino et al.~\cite{Anselmino:2011ch}, and other related studies.

Piloñeta and Vladimirov~\cite{Piloneta:2025jjb} studied kinematic power corrections to SIDIS for all unpolarized and polarized structure functions; these corrections scale as $k_\perp/Q$. Ebert, Gao, and Stewart~\cite{Ebert:2021jhy} analyzed next-to-leading-power (NLP) contributions to SIDIS, including kinematic corrections as well as subleading contributions arising from hard scattering and insertions of the subleading $\mathrm{SCET}{\mathrm{II}}$ Lagrangian. They present NLP corrections to both unpolarized and polarized structure functions; in the unpolarized case, the leading contribution at this order appears in $F_{UU}^{\cos\phi_h}$. Anselmino et al.~\cite{Anselmino:2011ch} consider a simple parton model and retain exact kinematics of on-shell quarks at the level of quark scattering. Although an expansion is possible, the formulas in Ref.~\cite{Anselmino:2011ch} typically include only leading-power (LP) terms.

For the convenience of the reader, we list conversions between the notations used in these papers and our notation in Appendix~\ref{app:conv}.

Our approximation consists of retaining the terms from $L^{\mu\nu}W^1_{\mu\nu}$ that have proven useful for the Drell–Yan process (see Ref.~\cite{Balitsky:2021fer}). Further improvements could be achieved by including contributions from $L^{\mu\nu}W^2_{\mu\nu}$ and $L^{\mu\nu}W^3_{\mu\nu}$. We do not perform a complete phenomenological analysis or a full extraction of TMDs; instead, we present estimates that illustrate the impact of next-to-leading-power corrections in semi-inclusive deep inelastic scattering.

For the simplicity of the numerical estimates, we employ the generalized parton model for TMDs. This framework has been shown to provide a reasonable description of multiplicities and asymmetries observed in SIDIS~\cite{Anselmino:2005nn,Collins:2005ie,DAlesio:2007bjf,Barone:2008tn,Schweitzer:2010tt, Signori:2013mda,Anselmino:2013lza,Barone:2015ksa,Cammarota:2020qcw,Gamberg:2022kdb}, although it lacks certain theoretical refinements, such as a complete treatment of TMD evolution. Since our goal is to estimate power corrections, the logarithmic dependence associated with evolution is not expected to play a significant role at this stage.

We will use the following parametrizations for TMDs
\begin{subequations}\ba
	f^a_1(x,{\bm k_\perp^2}) &=& f^a_1(x)\;
    	\frac{1}{\,\pi\avkperp_{f_1}}\;e^{-{\bm k_\perp^2}/\avkperp_{f_1}} \, ,
	\label{eq:Gauss-f1}\\
    	D^a_1(\zh,{\bm p_\perp^2}) &=& D_1^a(\zh)\,
    	\frac{1}{\,\pi\avpperp_{D_1}}\;e^{-{\bm p_\perp^2}/\avpperp_{D_1}} \, ,
	\label{eq:Gauss-D1}\\
	H_{1}^{\perp a}(\zh,{\bm p_\perp^2}) &=&  H_{1}^{\perp (1) a}(\zh) \;
	\frac{2 z^2 \mh^2}{\pi \avpperp_{H_{1}^\perp}^2} \;
	e^{-{\bm p_\perp^2}/{\avpperp_{H_{1}^\perp}}}\, ,\label{eq:Gauss-H1perp}\\
	h_{1}^{\perp a}(x,{\bm k_\perp^2}) &=&  h_{1}^{\perp (1) a}(x)\;
   	\frac{2 M^2}{\pi \avkperp_{h_{1}^\perp}^2}\;
 	e^{-{\bm k_\perp^2}/{\avkperp_{h_{1}^\perp}}}\,
	\label{eq:Gauss-h1perp}\, .
\ea\end{subequations}
These parametrizations have been shown to work reasonably well in phenomenological applications. The collinear functions $f_1^a(x)$ and $D_1^a(z)$ are taken from the MSTW~\cite{Martin:2009iq} and DSS~\cite{deFlorian:2007aj} extractions~\cite{Anselmino:2005nn,Collins:2005ie,DAlesio:2007bjf,Barone:2008tn,Schweitzer:2010tt, Signori:2013mda,Anselmino:2013lza,Barone:2015ksa,Cammarota:2020qcw,Gamberg:2022kdb}. This setup is the same as in Ref.~\cite{Bastami:2018xqd}, where semi-inclusive deep inelastic scattering was studied within the Wandzura–Wilczek-type (WW) approximation.

In this approximation, TMD functions that originate from quark–gluon correlations are assumed to be small, and a series of relations among TMDs can be established. These relations, such as $x f^{\perp} \simeq f_1$ and $x h \simeq -({\bm k_\perp^2}/{M^2}) h_1^\perp$ (see Appendix~\ref{app:misc}), allow one to significantly reduce the number of TMDs contributing to structure functions and to simplify the expressions. It was found in Ref.~\cite{Bastami:2018xqd} that the WW approximation is generally in agreement with experimental observations. By adopting the same setup, we will be able to comment on the similarities and differences between the WW approximation and our formalism.

Parametrizations for unpolarized TMD PDF and TMD FF will be taken from Ref.~\cite{Anselmino:2005nn}, they are flavor independent with $\avkperp_{f_1} = 0.25$ (GeV$^2$), $\avpperp_{D_1}= 0.2$ (GeV$^2$). As we do not perform a phenomenological analysis, these parametrizations are sufficient for numerical estimates. Using formulas ~\eqref{eq:FUU_Ian1}-\eqref{eq:FUUcos2phi_Ian1}
one can perform calculations with any other set of TMD PDFs and TMD FFs.
Boer-Mulders functions and their first moments, $h_1^{\perp (1)} (x)$, are taken from Ref.~\cite{Barone:2009hw} and the Collins FF and their first moments, $H_1^{\perp (1)} (\zh)$, are from  Ref.~\cite{Anselmino:2013vqa}. The initial scale of the parametrizations in Eqs.~(\ref{eq:Gauss-f1} - \ref{eq:Gauss-h1perp}) is $Q^2=2.4$ GeV$^2$. This setup will allow us to have the direct comparison with results of Ref.~\cite{Bastami:2018xqd}. In the plots in the following Sections we will not restrict the range of $\Phperp$ to the one where TMD formalism works, $q_T/Q \ll 1$. The reason is that, first, experimental values of $q_T$ are not always available and, second, we would like to explore the expression for the whole range in $\Phperp$; however the interpretation of our estimates at large values of $\Phperp$ should be done with caution.

Notice that it was found in Ref.~\cite{Barone:2009hw} that both $u$ and $d$ first moments of Boer-Mulders functions are negative in SIDIS. Collins FF for $u$ quark fragmentation into $\pi^+$ is positive, while $d$ quark fragmentation into $\pi^+$ is negative, see e.q. Ref.~\cite{Anselmino:2013vqa}.

\subsection{$F_{UU,T}$ structure function}
\label{sec:fuut}
Our result for $F_{UU,T}$, Eq.~\eqref{eq:FUU_Ian1}, shows that it has LP and NNLP contributions, while NLP contribution is absent, 
\begin{align} 
 F_{UU,T}	&=  F_{UU,T}^{\rm LP} + F_{UU,T}^{\rm NNLP} \; ,  \\
 F_{UU,T} & =  {\cal C} \left[ f_1  D_1  \right]   
   + {\cal C}\biggl[ \frac{2 q_T }{Q^2} (\bfhp\cdot {\bm k_\perp}) f_1  D_1\biggr]  +{\cal C}\biggl[\frac{2 q_T {\bm k_\perp^2}}{\zh M m_h Q^2} (\bfhp \cdot {\bm p_\perp})  \kappa h_1^\perp  H_1^\perp \biggr] 
\,, \label{eq:FUUT-final}
\end{align}
where the first term in \eq{FUUT-final} contains the standard leading power result, 
\begin{align} 
F_{UU,T}^{\rm LP}	&= {\cal C} \left[ f_1 D_1 \right] \; ,
\end{align}
see Refs~\cite{Bacchetta:2006tn,Boussarie:2023izj}. The next-to-next-to-leading power contributions $F_{UU,T}^{\rm NNLP}$, the second and the third terms in \eq{FUUT-final} proportional to $1/Q^{2}$, are the main results of our study. This formula appears to have a slightly different structure comparing to Ref.~\cite{Piloneta:2025jjb}, see Eq.~(B.3) from Ref.~\cite{Piloneta:2025jjb}:
\begin{align} 
F_{UU,T}|_{\mbox{\tiny Ref.\cite{Piloneta:2025jjb}}} = {\cal C} \left[ f_1 D_1 \right] + \frac{1}{2} F_{UU,L} \; ,
\end{align}
where $F_{UU,L}$ is discussed in the next Section.
The difference of our result and Ref.~\cite{Piloneta:2025jjb} is due to different results for $\mathcal{S}_0^{\mu\nu} W_{\mu\nu}$ in the two formalisms.

In order to provide numerical estimates we will use parametrizations from \eqs{Gauss-f1}{Gauss-D1}. We obtain the following expressions
\begin{align}
	F_{UU,T}^{\rm LP}(x,\zh,\Phperp)
	&= x \sum_a e_a^2\,f^a_1(x)\,D_1^a(\zh)\,\frac{\exp(-\Phperp^2/(\zh^2\,\avkperp_{f_1} + \avpperp_{D_1}))}{\pi\,(\zh^2\,\avkperp_{f_1} + \avpperp_{D_1})}\,,
	\label{eq:FUUT-LP-gauss}
\end{align}
and
\begin{align}
	F_{UU,T}^{\rm NNLP}(x,\zh,\Phperp)
	&= x \sum_a e_a^2\,f^a_1(x)\,D_1^a(\zh)\,\frac{2 \avkperp_{f_1}\Phperp^2}{Q^2}\frac{\exp(-\Phperp^2/(\zh^2\,\avkperp_{f_1} + \avpperp_{D_1}))}{\pi\,(\zh^2\,\avkperp_{f_1} + \avpperp_{D_1})^2}\, \nonumber  \\
    &+x \sum_a e_a^2\,h_{1}^{\perp (1)a}(x)\,H_1^{\perp(1)a}(\zh) \frac{ \lambda \Phperp^2 }{Q^2} \frac{\exp(-\Phperp^2/(\zh^2\,\avkperp_{h_1^\perp} + \avpperp_{H_1^\perp}))}{\pi\,(\zh^2\,\avkperp_{h_1^\perp} + \avpperp_{H_1^\perp})^4}\,,
	\label{eq:FUUT-NLP-gauss}
\end{align}
where $\lambda = 8 M\mh (\avpperp_{H_1^\perp}^2 +\zh^2 \avkperp_{h_1^\perp}(\Phperp^2 -\zh^2 \avkperp_{h_1^\perp}))$.

\begin{figure}[h]
\centering
  \includegraphics[width=0.461\textwidth]{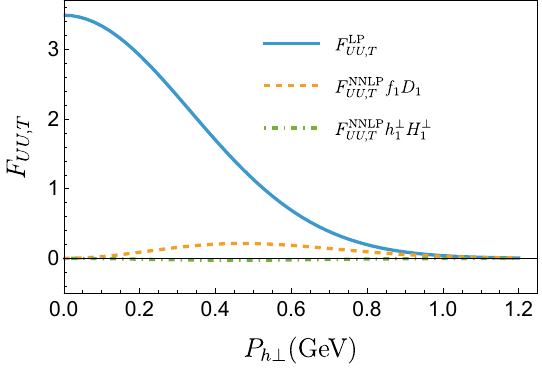}
  \includegraphics[width=0.5\textwidth]{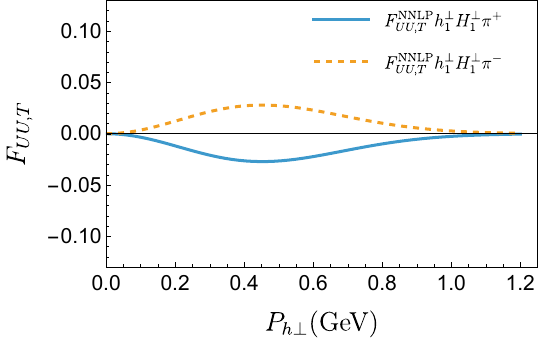}
\caption{\label{fig:FUUT} 
 $F_{UU,T}$  structure function for pion production off the proton at $Q^2=3$ GeV$^2$, $x=0.2$, $\zh = 0.3$.  Left panel shows $\pi^+$ leading power contribution (blue line), next-to-next-to-leading power contribution from $f_1 D_1$ (orange dashed line), and from $h_1^\perp H_1^\perp$ (green dot-dashed line).  Right panel shows next-to-next-to-leading power contribution from $h_1^\perp H_1^\perp$ for $\pi^+$ (blue line) and $\pi^-$ (orange dashed line) production.}
\end{figure}

In \fig{FUUT} we show various contributions fo $F_{UU,T}$ for the pion production off the proton at characteristic values for the kinematical variables $Q^2=3$ GeV$^2$, $x=0.2$, $\zh = 0.3$. One can see that $F_{UU,T}^{\rm NNLP}$ is dominated by the contribution from $f_1 D_1$, second term in  \eq{FUUT-final}. Contributions from $h_1^\perp H_1^\perp$, third term in \eq{FUUT-final}, have opposite signs and approximately the same magnitudes for $\pi^+$ and $\pi^-$, see the left panel of \fig{FUUT}. We will show the comparison to experimental data in the following Section.

\begin{figure}[h]
\centering
  \includegraphics[width=0.55\textwidth]{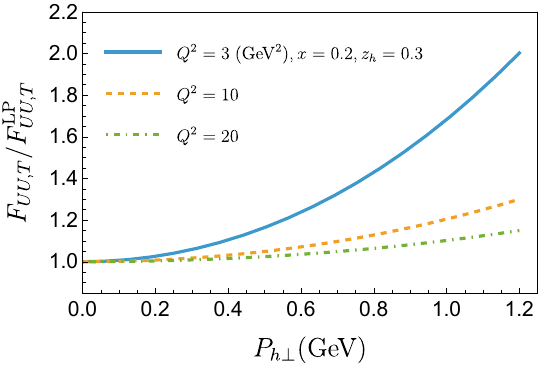}
\caption{\label{fig:FUUT_ratio} 
 The ratio of $F_{UU,T}$  structure function over the leading power $F_{UU,T}^{\rm LP}$ for $\pi^+$ production off the proton at $Q^2=3$ GeV$^2$, $x=0.2$, $\zh = 0.3$.}
\end{figure}

The ratio $F_{UU,T}$  structure function over the leading power $F_{UU,T}^{\rm LP}$ for $\pi^+$ production off the proton at three values for $Q^2$ and  $x=0.2$, $\zh = 0.3$ is shown in \fig{FUUT_ratio}. One can see that the next-to-next-to-leading power contribution becomes large at large values of $\Phperp$, even surpassing the leading power contribution at large $\Phperp$ for relatively low values of $Q^2$. Even for the largest $Q^2=20$ GeV$^2$ shown in \fig{FUUT_ratio}, the ratio can be around 10\%. As experiments often operate with low to moderate values of $Q^2$ we conclude that it is important to take into account next-to-next-to-leading power contributions for $F_{UU,T}$ in phenomenological analyses of the experimental data. In addition, taking into account $F_{UU,T}^{\rm NNLP}$ will be important for transition region from TMD to collinear description of the experimental data that will happen in the region of $q_T \sim Q$.

\subsection{$F_{UU,L}$  structure function}
\label{sec:fuul}

Our result for $F_{UU,L}$  structure function, that starts from next-to-next-to-leading power, Eq.~\eqref{eq:FUUL_Ian1}, reads
\begin{align}
    F_{UU,L}	&= {\cal C} \biggl[  \frac{4 {\bm k_\perp^2}}{Q^2} f_1 D_1\biggr] + {\cal C} \biggl[\frac{4 {\bm k_\perp^2}}{\zh M m_h Q^2} ({\bm k_\perp} \cdot {\bm p_\perp}) \kappa h_1^\perp H_1^\perp \biggr]
    \label{eq:FUUL}
\,.
\end{align}
It is the same as the result in Pilo\~neta and Vladimirov, Ref.~\cite{Piloneta:2025jjb}, see Eq.~(B.4) from Ref.~\cite{Piloneta:2025jjb}, calculated for large $Q^2$. $F_{UU,L}$ appears only at next-to-next-to-leading power and therefore was not used previously in TMD phenomenology. Nevertheless, the first term in \eq{FUUL} was known and calculated in Bacchetta et al Ref.~\cite{Bacchetta:2008xw}, see Eq.(6.15) of Ref.~\cite{Bacchetta:2008xw}. It was obtained by employing the generalized parton model of Anselmino et al~\cite{Anselmino:2005nn}. In Ref.~\cite{Anselmino:2005nn} the intrinsic transverse momentum is included in distribution and fragmentation functions and the kinematics is taken such that the quarks in the parton sub-process of quark scattering are on the mass shell. It is remarkable, but not surprising, that the parton model considered in Ref.~\cite{Anselmino:2005nn} gives the correct results for the kinematical next-to-next-to-leading power contribution to $F_{UU,L}$ and, as we will see in the following sections, also for $F_{UU}^{\cos 2\phi_h}$ and $F_{UU}^{\cos \phi_h}$. Notice that  Ref.~\cite{Anselmino:2005nn} was later developed as generalized helicity formalism by Anselmino et al in Ref.~\cite{Anselmino:2011ch} and was shown to coincide in kinematical contributions with the standard leading power results, such as Ref.~\cite{Bacchetta:2006tn}. 

There exists a considerable experimental interest to study $F_{UU,L}$. The ratio   
\begin{align}
R_{\rm SIDIS} =  \frac{F_{UU,L}}{F_{UU,T}}
\end{align}
is planned to be studied at Jefferson Lab by experiments {\tt E12-06-104, E12-09-017, E12-09-002} in Hall C and Hall B, and the first data from HALL C have already appeared in Ref.~\cite{HallCSIDIS:2025plr}.

We obtain the following parametrization for $F_{UU,L}$:
\begin{align}
	F_{UU,L}(x,\zh,\Phperp)
	&= x \sum_a e_a^2\,f^a_1(x)\,D_1^a(\zh)\,\frac{\lambda_1}{Q^2}\frac{\exp(-\Phperp^2/(\zh^2\,\avkperp_{f_1} + \avpperp_{D_1}))}{\pi\,(\zh^2\,\avkperp_{f_1} + \avpperp_{D_1})^3}\, \nonumber  \\
    &-x \sum_a e_a^2\,h_{1}^{\perp (1)a}(x)\,H_1^{\perp(1)a}(\zh) \frac{ \lambda_2  }{Q^2} \frac{\exp(-\Phperp^2/(\zh^2\,\avkperp_{h_1^\perp} + \avpperp_{H_1^\perp}))}{\pi\,(\zh^2\,\avkperp_{h_1^\perp} + \avpperp_{H_1^\perp})^5}\,,
	\label{eq:FUUL-gauss}
\end{align}
where $\lambda_1 = 4 \avkperp_{f_1} (\avpperp_{D_1}^2 +\zh^2 \avkperp_{f_1}(\Phperp^2 + \avpperp_{D_1}))$ and $\lambda_2 = 16 M\mh \zh^2 \avkperp_{h_1^\perp} (-\avkperp_{h_1^\perp}\Phperp^4\zh^2 + 2 \avpperp_{H_1^\perp} (\avpperp_{H_1^\perp}+\zh^2 \avkperp_{h_1^\perp})^2 - 2 \Phperp^2(\avpperp_{H_1^\perp}^2-\zh^4 \avkperp_{h_1^\perp}^2) )$.

\begin{figure}[h]
\centering
  \includegraphics[width=0.49\textwidth]{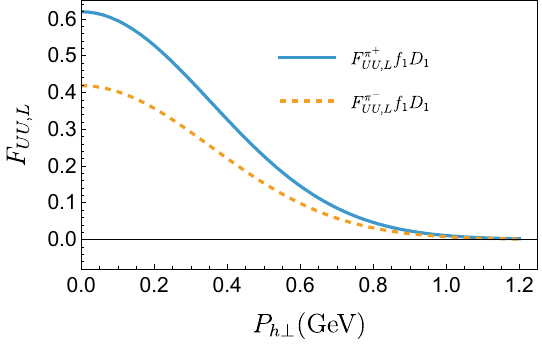}
  \includegraphics[width=0.49\textwidth]{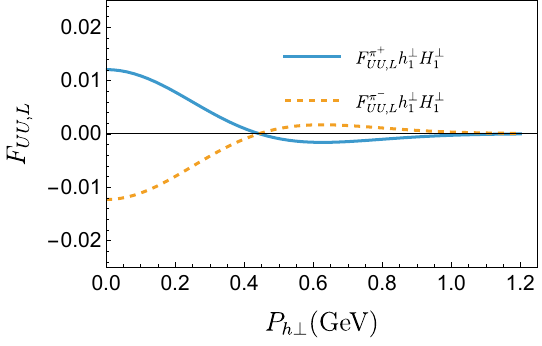}
\caption{\label{fig:FUUL} 
 $F_{UU,L}$  for $\pi^+$  and $\pi^-$ production off the proton at $Q^2=5$ GeV$^2$ and $x=0.2$, $\zh = 0.3$. The contribution from the first term in \eq{FUUL}, convolution of $f_1 D_1$,  is shown in the left panel as blue line for $\pi^+$ and orange dashed line for $\pi^-$. The second term, convolution of $h_1^\perp D_1^\perp$, in \eq{FUUL} is shown in the right panel as blue line for $\pi^+$ and orange dashed line for $\pi^-$.}
\end{figure}

Our numerical estimates of $F_{UU,L}$ for $\pi^+$ and $\pi^-$  production off the proton at $Q^2=5$ GeV$^2$ at $x=0.2$, $\zh = 0.3$ are shown in \fig{FUUL}. One can see that the contribution from the first term in \eq{FUUL} is always positive, it is around 20\% of $F_{UU,T}$ at this $Q^2$. The second term, proportional to $h_1^\perp D_1^\perp$, changes sign between $\pi^+$ and $\pi^-$. It means that there may be differences in experimental results for $\pi^+$ and $\pi^-$ production. Notice that the term proportional to $h_1^\perp D_1^\perp$ gives zero if integrated over $\Phperp$, therefore we expect to observe these differences in $\Phperp$ dependent measurements only.

\begin{figure}[h]
\centering
  \includegraphics[width=0.49\textwidth]{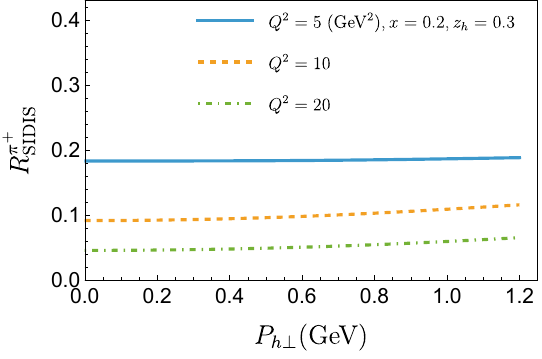}
  \includegraphics[width=0.49\textwidth]{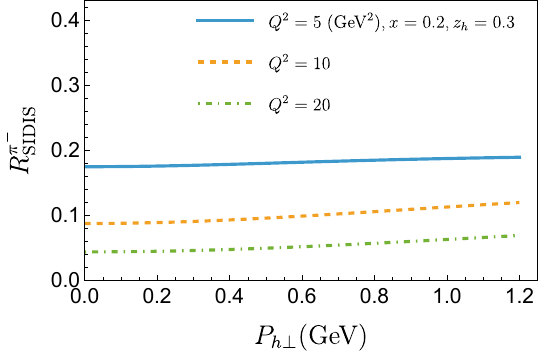}
\caption{\label{fig:RSIDIS} 
 $R_{\rm SIDIS}$  for $\pi^+$ (left panel) and $\pi^-$ (right panel) production off the proton at $Q^2=5,10,20$ GeV$^2$ and $x=0.2$, $\zh = 0.3$.}
\end{figure}

We show our estimates for $R_{\rm SIDIS}$ in \fig{RSIDIS} where we observe that the values are slightly different for $\pi^+$ and $\pi^-$ production due to the term containing $h_1^\perp H_1^\perp$ that has opposite sign for $\pi^+$ and $\pi^-$. We predict this ratio to be around 20\% in the kinematical range of Jefferson Lab. $R_{\rm SIDIS}$ can also be studied at the Electron-Ion Collider at higher values of $Q$. The Electron-Ion Collider data will allow to study the transition region from low to high $q_T$. Our estimates are in general agreement with those presented in Ref.~\cite{Accardi:2023chb}.

\begin{figure}[h]
\centering
  \includegraphics[width=0.49\textwidth]{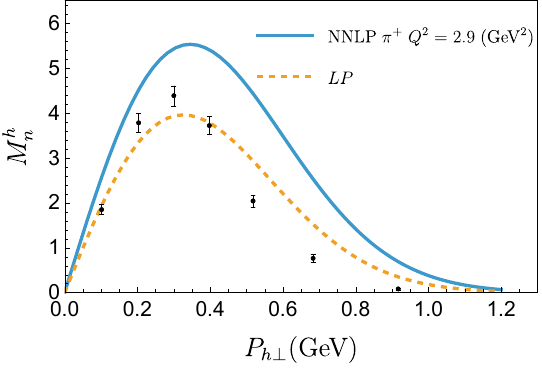}
  \includegraphics[width=0.49\textwidth]{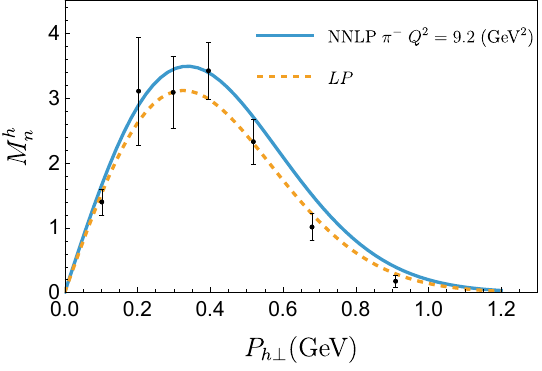}
\caption{\label{fig:HERMES}  SIDIS multiplicity measured by HERMES~\cite{HERMES:2012uyd} for $\pi^+$ (right panel)and $\pi^-$ (left panel) production off the proton target. Calculations shown as blue lines include NNLP contributions, the orange dashed lines include only leading power terms.}
\end{figure}

At this point it is interesting to compare our results to the existing experimental data. HERMES experiment measured multiplicities for pion and kaon production in the scattering of 27.6 GeV
electrons or positrons of HERA's polarized lepton storage ring off proton and deuteron targets. 
The HERMES multiplicity is defined as~\cite{HERMES:2012uyd}
\be\label{eq:multiplicity-HERMES}
	M_n^h(x,\zh,\Phperp) \equiv
	\frac{d\sigma_{\rm SIDIS}(x,\zh,\Phperp)/dx\,d\zh\,d\Phperp}
	{d\sigma_{\rm DIS}(x)/dx} =
	2 \pi \Phperp \frac{F_{UU,T}+p_1 F_{UU,L}}{ x \sum_q e_q^2\,f^q_1(x)}\, .
\ee
In this formula, the second term $p_1 F_{UU,L}$ was not taken into account in the TMD related studies. We present our numerical estimates for two characteristic bins with moderate and high $Q^2$ in 
\fig{HERMES} on the proton target~\cite{HERMES:2012uyd}, the left panel shows the estimate for $\pi^+$ production at $\la Q^2\ra=2.87\,({\rm GeV}^2)$, $\la x\ra=0.15$, $\la z\ra=0.22$, while the right panel shows the estimate for $\pi^-$ production at $\la Q^2\ra=9.2\,({\rm GeV}^2)$, $\la x\ra=0.41$, $\la z\ra=0.22$. The estimates including $F_{UU,T}$ and  $F_{UU,L}$ at to next-to-next-to-leading power are shown in blue, while the calculations including only $F_{UU,T}$ at leading power are shown as orange dashed lines. One can see that at moderate values of $Q^2$ the NNLP contributions are substantial and would influence the phenomenological results. It demonstrates the importance of accounting for NNLP contributions in the future studies. These contributions may be important in resolution of the normalization puzzle found in SIDIS in Ref.~\cite{Bacchetta:2022awv}. Notice that the collinear cross section $d\sigma_{\rm DIS}/dx$ also contains contributions due to the longitudinally polarized photons that will be needed to be taken into account in phenomenological studies, we neglect this contribution here for simplicity.

Finally, we show two representative plots from the COMPASS experiment
where charged pions, kaons, or hadrons were measured with 160 GeV
longitudinally polarized muons scattered off proton and deuteron
targets. The multiplicity measured by COMPASS reads
\begin{align}
\label{eq:multiplicity-COMPASS}
	n^h(x,\zh,\Phperp^2)  \equiv
	\frac{d\sigma_{\rm SIDIS}(x,z,\Phperp^2)/dx\,dz\,d\Phperp^2}
	{d\sigma_{\rm DIS}(x)/dx} =
	\pi \frac{F_{UU,T}+p_1 F_{UU,L}}{ x \sum_q e_q^2\,f^q_1(x)}\;.
\end{align}

\begin{figure}[h]
\centering
  \includegraphics[width=0.49\textwidth]{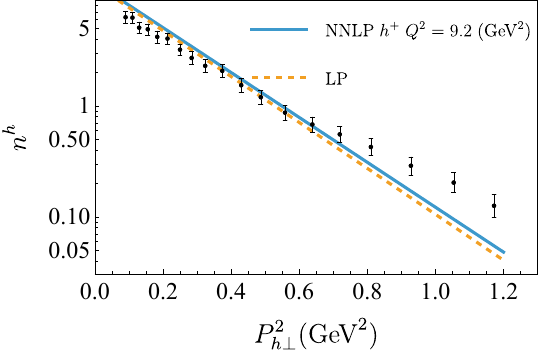}
  \includegraphics[width=0.49\textwidth]{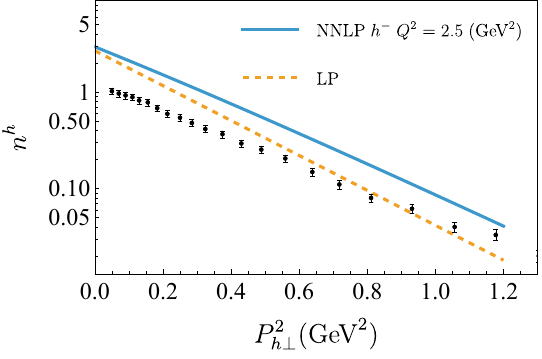}
\caption{\label{fig:COMPASS}  SIDIS multiplicity measured by COMPASS~\cite{COMPASS:2017mvk} for $h^+$ (left panel)and $h^-$ (right panel) production off the deuterium target. Calculations shown as blue lines include NNLP contributions, the orange dashed lines include only leading power terms.}
\end{figure}

Left panel of \fig{COMPASS} shows the COMPASS multiplicity~\cite{COMPASS:2017mvk}
at $\la Q^2\ra=20\,{\rm GeV}^2$, $\la x\ra  =0.15$, $\la z\ra  =0.2$
for $h^+$ production on the deuterium target. Right panel of \fig{COMPASS} shows the COMPASS multiplicity~\cite{COMPASS:2017mvk}
at $\la Q^2\ra=2.5\,{\rm GeV}^2$, $\la x\ra  =0.05$, $\la z\ra  =0.4$
for $h^-$ production on the deuterium target. One can see that even at COMPASS energy NNLP contributions are substantial, especially for low values of $Q^2$. In order to calculate the multiplicity we assumed that charged hadrons are dominated by pions and we used the same parameters for TMD PDF and TMD FF as those we used for calculations for HERMES. One can see that our simple model does not capture well the structure of the COMPASS data.

\begin{figure}[h]
\centering
  \includegraphics[width=0.55\textwidth]{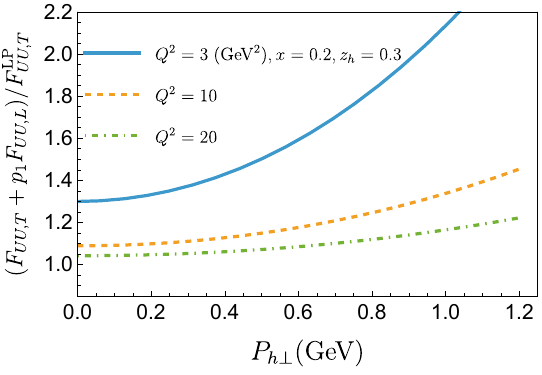}
\caption{\label{fig:FUUTFUUL_ratio} 
 The ratio of $F_{UU,T}+p_1 F_{UU,L}$  structure functions over the leading power $F_{UU,T}^{\rm LP}$ for $\pi^+$ production off the proton at $Q^2=3$ GeV$^2$, $x=0.2$, $\zh = 0.3$.}
\end{figure}

We conclude at this point that taking into account the longitudinal structure function $F_{UU,L}$ is important in phenomenological applications.
In order to further demonstrate it,  in \fig{FUUTFUUL_ratio} we plot the ratio $F_{UU,T}+p_1 F_{UU,L}$  structure functions over the leading power $F_{UU,T}^{\rm LP}$ for $\pi^+$ production off the proton at $Q^2=3$ GeV$^2$, $x=0.2$, $\zh = 0.3$ . One can see, compare to \fig{FUUT_ratio}, that the $F_{UU,L}$ is not negligible even at small  values of $\Phperp$.

\subsection{$F_{UU}^{\cos\phi_h}$  structure function}
\label{sec:fuucosphi}

$F_{UU}^{\cos\phi_h}$  structure function starts from the next-to-leading contribution ($\sim1/Q$)~\footnote{The next contribution to $F_{UU}^{\cos\phi_h}$ will be at $1/Q^3$, NNNLP.}. It appears to have many various kinematical and dynamical contributions \cite{Bacchetta:2006tn}. We consider contributions from $f_1 D_1$ and $h_1^\perp H_1^\perp$, see  Eq.~\eqref{eq:FUUcosphi_Ian1}, 
\begin{align}
    F_{UU}^{\cos\phi_h}
	&=
	{\cal C}\biggl[ -\frac{2 (\bfhp \cdot{\bm k_\perp})}{Q} f_1 D_1\biggr] +{\cal C}\biggl[ - \frac{2{\bm k_\perp^2}}{\zh M m_h Q} ({\bfhp}\cdot{\bm p_\perp})
	\kappa h_1^\perp H_{1}^{\perp }
	\biggr] , \label{eq:FUUcosphi-final}
\end{align}
This result coincides with that of Piloñeta and Vladimirov~\cite{Piloneta:2025jjb} (see Eq.~(B.5) therein), as well as with the result of Ebert, Gao, and Stewart~\cite{Ebert:2021jhy}, after neglecting the “tilde” functions in Eq.~(5.41). It also agrees with the expression given by Anselmino et al.~\cite{Anselmino:2011ch} (see Eq.~(65) of that reference). Furthermore, the formula matches the result of Bacchetta et al.~\cite{Bacchetta:2006tn} upon applying the equations of motion (EOMs) and neglecting the “tilde” functions in Eq.~(4.4); see Appendix~\ref{app:misc} for details.

Historically, the $\cos \phi_h$ modulation in the SIDIS cross section was proposed in the 1970s by R.~Cahn~\cite{Cahn:1978se} as evidence for the intrinsic transverse momentum of quarks inside the nucleon. The first term in \eq{FUUcosphi-final} is therefore known as the Cahn term, as it arises kinematically from the transverse motion of quarks.

We obtain the following parametrization for $F_{UU}^{\cos\phi_h}$:
\begin{align}
	F_{UU}^{\cos\phi_h}(x,\zh,\Phperp)
	&= -x \sum_a e_a^2\,f^a_1(x)\,D_1^a(\zh)\,\frac{2 \avkperp_{f_1} \Phperp \zh}{Q}\frac{\exp(-\Phperp^2/(\zh^2\,\avkperp_{f_1} + \avpperp_{D_1}))}{\pi\,(\zh^2\,\avkperp_{f_1} + \avpperp_{D_1})^2}\, \nonumber  \\
    &-x \sum_a e_a^2\,h_{1}^{\perp (1)a}(x)\,H_1^{\perp(1)a}(\zh) \frac{ \lambda_3  }{Q} \frac{\exp(-\Phperp^2/(\zh^2\,\avkperp_{h_1^\perp} + \avpperp_{H_1^\perp}))}{\pi\,(\zh^2\,\avkperp_{h_1^\perp} + \avpperp_{H_1^\perp})^4}\,,
	\label{eq:FUUcosphi-gauss}
\end{align}
where  $\lambda_3 = 8 M\mh \zh \Phperp (\avpperp_{H_1^\perp}^2 + \zh^2\,\avkperp_{h_1^\perp} (\Phperp^2 -\zh^2\,\avkperp_{h_1^\perp} ) )$.

\begin{figure}[h]
\centering
  \includegraphics[width=0.49\textwidth]{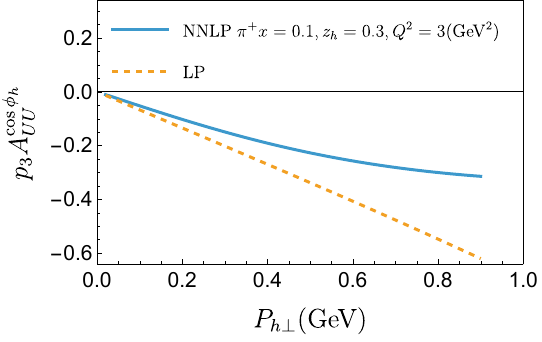}
  \includegraphics[width=0.49\textwidth]{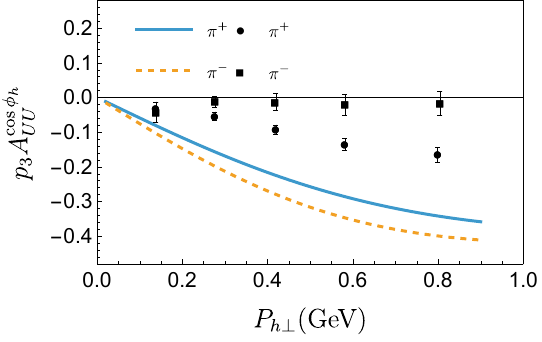}
\caption{\label{fig:Acosphi_HERMES} 
 Left panel, 
 $p_3 A_{UU}^{\cos \phi_h}$ for $\pi^+$ asymmetry calculated at characteristic HERMES kinematics,  $Q^2=3$ GeV$^2$, $x=0.1$, $\zh = 0.3$ r with NNLP terms (blue line) and with LP terms only (orange dashed line) in the denominator. Right panel shows the comparison of our estimates to HERMES experimental data \cite{HERMES:2012kpt} for $\pi^+$ (circles) and $\pi^-$ (squares).}
\end{figure}

The asymmetries $A_{UU}^{\cos\phi_h}\sim F_{UU}^{\cos\phi_h}/(F_{UU,T}+p_1 F_{UU,L})$ were measured by
EMC \cite{EuropeanMuon:1983tsy}, Jefferson Lab \cite{CLAS:2008nzy,Mkrtchyan:2007sr},
HERMES \cite{HERMES:2012kpt}, and COMPASS \cite{COMPASS:2014kcy}. In~\fig{Acosphi_HERMES} we compare our estimates to the HERMES experiment that measured azimuthal asymmetries~\cite{HERMES:2012kpt} defined as
\begin{align}
2 \langle \cos \phi_h \rangle \equiv p_3 A_{UU}^{\cos \phi_h} =  p_3 \frac{F_{UU}^{\cos\phi_h}}{F_{UU,T}+p_1 F_{UU,L}}\; .
\end{align}
In the left panel of \fig{Acosphi_HERMES} we plot $p_3 A_{UU}^{\cos \phi_h}$ for $\pi^+$ production on the hydrogen target at characteristic kinematics of HERMES, $x=0.1$, $\zh = 0.3$, and $Q^2 = 3$ (GeV$^2$), using NNLP terms (blue line) and using only leading power terms (dashed orange line) in the denominator of the asymmetry. Right panel of  \fig{Acosphi_HERMES} shows comparison of $p_3 A_{UU}^{\cos \phi_h}$ to the data for $\pi^+$ and $\pi^-$ production. The simple gaussian model that we use has difficulty in describing the data (especially for $\pi^-$ production), however the description is improved when NNLP terms are taken into account.

The COMPASS Collaboration measured $\cos \phi_h$ asymmetries in Ref.~\cite{COMPASS:2014kcy} defined as 
\begin{align}
A^{UU}_{\cos \phi_h} \equiv A_{UU}^{\cos \phi_h} =  \frac{F_{UU}^{\cos\phi_h}}{F_{UU,T}+p_1 F_{UU,L}}\;,
\end{align}
where the measurements were performed for $h^\pm$ hadrons produced in the scattering of 160~GeV muons off a deuterium target.

\begin{figure}[h]
\centering
  \includegraphics[width=0.485\textwidth]{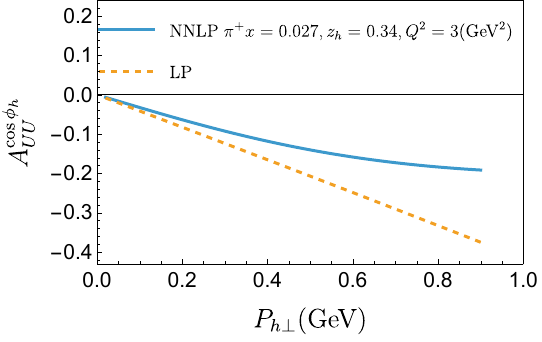}
  \includegraphics[width=0.485\textwidth]{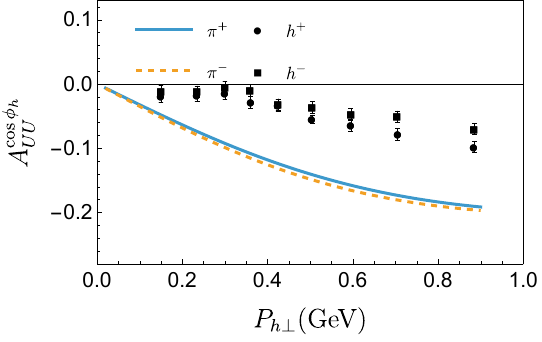}
\caption{\label{fig:Acosphi_COMPASS} 
 Left panel, 
 $A_{UU}^{\cos \phi_h}$ for $\pi^+$ asymmetry calculated at characteristic COMPASS kinematics,  $Q^2=3$ GeV$^2$, $x=0.027$, $\zh = 0.34$ with NNLP terms (blue line) and with LP terms only (orange dashed line) in the denominator. Right panel shows the comparison of our estimates for $A_{UU}^{\cos \phi_h}$ to COMPASS experimental data \cite{COMPASS:2014kcy} for $h^+$ (circles) and $h^-$ (squares).}
\end{figure}

In~\fig{Acosphi_COMPASS}, we compare our estimates with the COMPASS data~\cite{COMPASS:2014kcy}. Note that the COMPASS Collaboration measures $h^\pm$ hadrons; here, we approximate charged hadrons as charged pions, $\pi^\pm$. One can see from~\fig{Acosphi_HERMES} and~\fig{Acosphi_COMPASS} that next-to-next-to-leading-power (NNLP) contributions in the denominator reduce the size of the asymmetry, bringing the predictions closer to the measured values.

It is well known that a phenomenological description of the $\cos \phi_h$ and $\cos 2\phi_h$ modulations is challenging (see, e.g., Ref.~\cite{Barone:2015ksa}). The asymmetry is dominated by the Cahn term (the first term in \eq{FUUcosphi-final}). It was shown in Ref.~\cite{Barone:2015ksa} that, within a simple parton model, the $\cos \phi_h$ asymmetry is highly sensitive to the widths of the transverse-momentum distributions, with significantly smaller values obtained compared to standard choices. In addition, Ref.~\cite{Barone:2015ksa} found that the contribution from the Boer–Mulders function is negligible.

Indeed, as seen in~\fig{Acosphi_HERMES} and~\fig{Acosphi_COMPASS}, the predicted difference between $\pi^+$ and $\pi^-$, primarily driven by the Boer–Mulders contribution (the second term in \eq{FUUcosphi-final}), exhibits trends opposite to those observed experimentally. We therefore expect that a phenomenological analysis incorporating next-to-next-to-leading-power (NNLP) corrections will improve the overall description.

\subsection{$F_{UU}^{\cos2\phi_h}$ structure function}
\label{sec:fuucos2phi}
$F_{UU}^{\cos2\phi_h}$  structure function, Eq.~\eqref{eq:FUUcos2phi_Ian1}, has the leading and next-to-next-to-leading terms
\begin{align}
F_{UU}^{\cos 2\phi_h} &=  F_{UU}^{\cos 2\phi_h \rm LP} + F_{UU}^{\cos 2\phi_h \rm NNLP} \; ,  \\
F_{UU}^{\cos 2\phi_h} &={\cal C}\biggl[\frac{2(\bfhp\cdot{\bm k_\perp})(\bfhp\cdot{\bm p_\perp})-({\bm k_\perp}\cdot{\bm p_\perp})}{\zh M m_h} \kappa h_1^\perp H_{1}^{\perp}
	\biggr]
 \nonumber \\
 &  
  + {\cal C}\biggl[ \frac{2q_T (\bfhp \cdot {\bm k_\perp})}{Q^2}f_1 D_1 \biggr]+{\cal C}\biggl[\frac{2 q_T {\bm k_\perp^2}\; }{\zh M m_h Q^2} (\bfhp\cdot{\bm p_\perp})
	  \kappa h_1^\perp H_{1}^{\perp}
	\biggr] 
\,, \label{eq:FUUcos2phi}
\end{align}
where the leading-power contribution (the first term in \eq{FUUcos2phi}) coincides with the standard result (see Ref.~\cite{Bacchetta:2006tn}). It originates from the Boer–Mulders function $h_1^\perp$ convoluted with the Collins fragmentation function $H_1^\perp$. The next-to-next-to-leading-power contributions differ from those of Piloñeta and Vladimirov~\cite{Piloneta:2025jjb} (compare with Eq.~(B.6) therein), due to the different expressions obtained for $\mathcal{S}_0^{\mu\nu} W_{\mu\nu}$ in the two formalisms.

In our notations, the result of Piloñeta and Vladimirov~\cite{Piloneta:2025jjb} (see Eq.~(B.6) of that reference) reads:
\begin{align} 
F_{UU}^{\cos 2\phi_h} |_{\mbox{\tiny Ref.\cite{Piloneta:2025jjb}}} &= {\cal C}\biggl[\frac{2(\bfhp\cdot{\bm k_\perp})(\bfhp\cdot{\bm p_\perp})-({\bm k_\perp}\cdot{\bm p_\perp})}{\zh M m_h} \kappa h_1^\perp H_{1}^{\perp}
	\biggr]
 \nonumber \\
& + {\cal C}\biggl[ \frac{4(\bfhp \cdot {\bm k_\perp})^2}{Q^2}f_1 D_1 \biggr] - \frac{1}{2} F_{UU,L} \; \nonumber \\
& =  F_{UU}^{\cos 2\phi_h \rm LP}  + \frac{4M^2}{Q^2} \; {\cal C}\left[\;\frac{2\,(\bfhp\cdot{\bm k_\perp})^2-{\bm k_\perp^2}}{2M^2}\,f_{1}\,D_{1}\;
	\right]  - {\cal C} \biggl[\frac{2 {\bm k_\perp^2}}{\zh M m_h Q^2} ({\bm k_\perp} \cdot {\bm p_\perp}) \kappa h_1^\perp H_1^\perp \biggr] \label{eq:FUU_cos2phi_AV}
\end{align}
where the leading-power term coincides with our calculation, while the NNLP terms differ. In particular, the contribution to $F_{UU}^{\cos 2\phi}$ from $f_1 D_1$ in \eq{FUU_cos2phi_AV} agrees with the result of Bacchetta et al. in Ref.~\cite{Bacchetta:2008xw} (see Eq.~(6.15) therein), which was obtained using the generalized parton model of Anselmino et al.~\cite{Anselmino:2005nn}.

This contribution, in this form, dates back to the pioneering work of R.~Cahn~\cite{Cahn:1978se} and has been used in phenomenological studies of $\cos 2\phi_h$ asymmetries by Barone, Prokudin, and Ma~\cite{Barone:2008tn}, as well as by Barone, Melis, and Prokudin~\cite{Barone:2009hw}. In the future, it will be interesting to investigate $\cos 2\phi_h$ modulations in more detail, explore the differences between the methodology of this paper and other frameworks, and compare the results with experimental data.

We obtain the following parametrization for $F_{UU}^{\cos2\phi_h \; \rm LP}$:
\begin{align}
	F_{UU}^{\cos2\phi_h\; \rm LP}(x,\zh,\Phperp) =
    &x \sum_a e_a^2\,h_{1}^{\perp (1)a}(x)\,H_1^{\perp(1)a}(\zh) \lambda_4 \frac{\exp(-\Phperp^2/(\zh^2\,\avkperp_{h_1^\perp} + \avpperp_{H_1^\perp}))}{\pi\,(\zh^2\,\avkperp_{h_1^\perp} + \avpperp_{H_1^\perp})^3}\,,
	\label{eq:FUUcos2phi-gauss}
\end{align}
where $\lambda_4 = { 4 M m_h \Phperp^2 \zh^2 }$, while NNLP contributions will be calculated using \eq{relation}.

\begin{figure}[h]
\centering
  \includegraphics[width=0.49\textwidth]{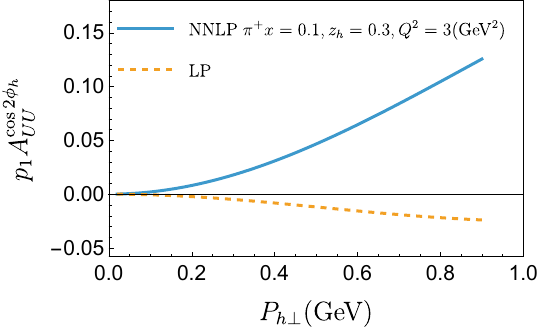}
  \includegraphics[width=0.49\textwidth]{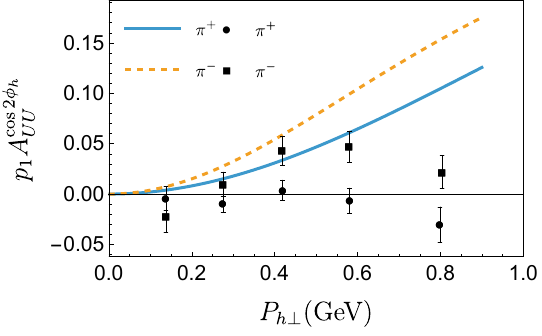}
\caption{\label{fig:Acos2phi_HERMES} 
 Left panel, 
 $p_1 A_{UU}^{\cos 2\phi_h}$ for $\pi^+$ asymmetry calculated at characteristic HERMES kinematics,  $Q^2=3$ GeV$^2$, $x=0.1$, $\zh = 0.3$ with NNLP terms (blue line) and with LP terms only (orange dashed line). Right panel shows the comparison of our estimates to HERMES experimental data \cite{HERMES:2012kpt} for $\pi^+$ (circles) and $\pi^-$ (squares).}
\end{figure}

In~\fig{Acos2phi_HERMES} we compare our estimates to the HERMES experiment that measured azimuthal asymmetries~\cite{HERMES:2012kpt} defined as
\begin{align}
2 \langle \cos 2\phi_h \rangle \equiv p_1 A_{UU}^{\cos 2\phi_h} =  p_1 \frac{F_{UU}^{\cos2\phi_h}}{F_{UU,T}+p_1 F_{UU,L}}\; .
\end{align}

In the left panel of \fig{Acos2phi_HERMES} we plot $p_1 A_{UU}^{\cos 2\phi_h}$ for $\pi^+$ production on the hydrogen target at characteristic kinematics of HERMES, $x=0.1$, $\zh = 0.3$, and $Q^2 = 3$ (GeV$^2$), using NNLP terms (blue line) and using only leading power terms (dashed orange line) in the denominator of the asymmetry. Right panel of  \fig{Acos2phi_HERMES} shows comparison of $p_1 A_{UU}^{\cos 2\phi_h}$ to the data for $\pi^+$ and $\pi^-$ production. 
Notice that in our formalism, the NNLP contributions to $F_{UU}^{\cos2\phi_h}$ are related to $F_{UU}^{\cos\phi_h}$, see \eq{relation}. One can see from the left panel of \fig{Acos2phi_HERMES} that the NNLP contributions change the result drastically. A large contribution from next-to-next-to-leading term in $\cos 2 \phi_h$ asymmetry was observed before in Refs.~\cite{Barone:2008tn,Barone:2009hw}.

\begin{figure}[h]
\centering
  \includegraphics[width=0.49\textwidth]{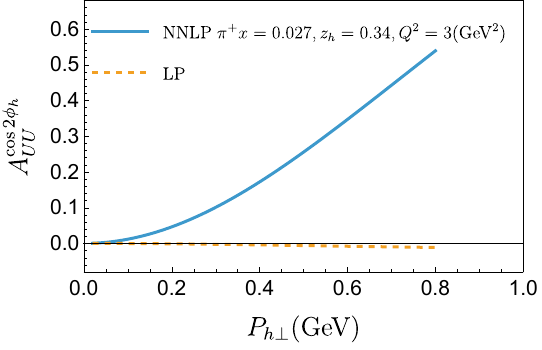}
  \includegraphics[width=0.49\textwidth]{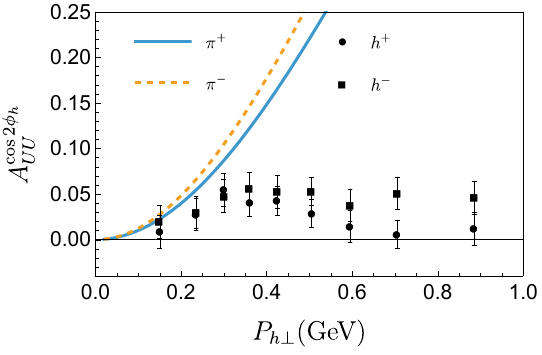}
\caption{\label{fig:Acos2phi_COMPASS} 
 Left panel, 
 $A_{UU}^{\cos 2\phi_h}$ for $\pi^+$ asymmetry calculated at characteristic COMPASS kinematics,  $Q^2=3$ GeV$^2$, $x=0.027$, $\zh = 0.34$ with NNLP terms (blue line) and with LP terms only (orange dashed line). Right panel shows the comparison of our estimates to COMPASS experimental data \cite{COMPASS:2014kcy} for $h^+$ (circles) and $h^-$ (squares) using the full calculation with leading and next-to-next-to-leading contributions.}
\end{figure}

COMPASS Collaboration measured the asymmetries in Ref.~\cite{COMPASS:2014kcy} defined as 
\begin{align}
A^{UU}_{\cos 2 \phi_h} \equiv A_{UU}^{\cos 2\phi_h} =  \frac{F_{UU}^{\cos2\phi_h}}{F_{UU,T}+p_1 F_{UU,L}}\;.
\end{align}
where the measurements are performed for $h^\pm$ hadrons produced in the scattering of muons at 160 GeV on deuterium target. In~\fig{Acos2phi_COMPASS} we compare our estimates for COMPASS data  \cite{COMPASS:2014kcy}. Notice that COMPASS collaboration measures $h^\pm$ while we approximate charged hadrons as charged pions $\pi^\pm$. One can see that the next-to-next-to-leading terms dominate the asymmetry and the trend is opposite to what is observed experimentally.

The Boer–Mulders functions used in our estimates, taken from Ref.~\cite{Barone:2009hw}, were extracted with the inclusion of the next-to-next-to-leading Cahn-type contribution (see \eq{FUUcos2phi_Bacchetta}). The extraction in Ref.~\cite{Barone:2009hw} encountered difficulties in describing the $\Phperp$-dependent data from HERMES and COMPASS. Similar issues were also observed in the analysis of Ref.~\cite{Barone:2015ksa}.

Based on our findings, we conclude that the leading-power Boer–Mulders contribution to $F_{UU}^{\cos2\phi_h}$ may not be the dominant one, and that a new phenomenological analysis is required to better understand the measured asymmetry. Indeed, as seen in the right panel of \fig{Acos2phi_HERMES}, the estimates that include NNLP contributions do not quantitatively describe the data; however, they exhibit similar trends, namely that the asymmetry for $\pi^-$ appears to be larger than that for $\pi^+$.

\begin{figure}[h]
\centering
\includegraphics[width=0.49\textwidth]{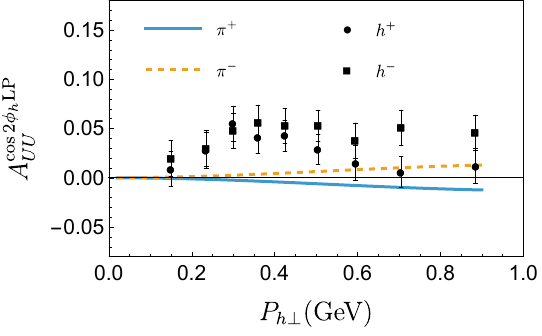}
  \includegraphics[width=0.49\textwidth]{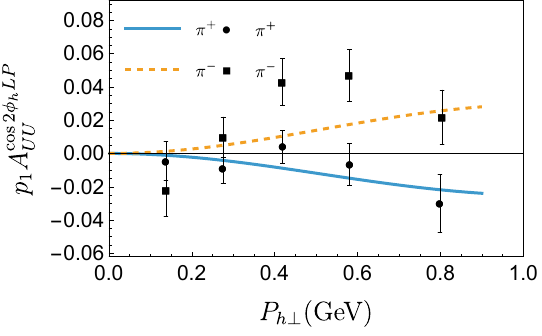}
\caption{\label{fig:Acos2phi_HERMES_LP} 
Left panel, 
 $A_{UU}^{\cos 2\phi_h \rm LP}$ for $\pi^+$ asymmetry with leading power terms for $\pi^+$ (blue line) and $\pi^-$ (orange dashed line) compared to COMPASS experimental data \cite{COMPASS:2014kcy} for $h^+$ (circles) and $h^-$ (squares).
 Right panel, 
 $p_1 A_{UU}^{\cos 2\phi_h}$ asymmetry with only leading power Boer-Mulders contribution  for $\pi^+$ (blue line) and $\pi^-$ (orange dashed line) compared to HERMES experimental data \cite{HERMES:2012kpt} for $\pi^+$ (circles) and $\pi^-$ (squares).}
\end{figure}

In order to further demonstrate the difficulty of the description of the data with leading power terms only, we plot in \fig{Acos2phi_HERMES_LP} the leading power terms only:
\begin{align}
 p_1 A_{UU}^{\cos 2\phi_h \rm LP} &=  p_1 \frac{F_{UU}^{\cos2\phi_h \rm LP}}{F_{UU,T}^{\rm LP}}\; , \nonumber \\
 A_{UU}^{\cos 2\phi_h \rm LP} &=  \frac{F_{UU}^{\cos2\phi_h \rm LP}}{F_{UU,T}^{\rm LP}}\; ,
\end{align}
for both COMPASS (left panel of \fig{Acos2phi_HERMES_LP}) and HERMES (right panel of \fig{Acos2phi_HERMES_LP}) experiments. If one assumes the dominance of pions in the charged hadrons, then it is clear that the data are not described by the leading power contributions, as the leading power contributions are likely to have opposite signs for positive and negative hadrons.

\subsection{Study of relations \eq{relation}}
\begin{figure}[h]
\centering
\includegraphics[width=0.49\textwidth]{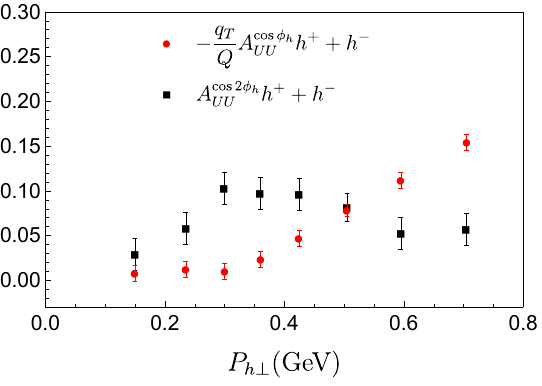}
\caption{\label{fig:relation} 
COMPASS experimental data \cite{COMPASS:2014kcy} manipulated as described in the text, $-q_T/Q(A_{UU}^{\cos \phi_h h^+} + A_{UU}^{\cos \phi_h h^-})$ red circles, $A_{UU}^{\cos 2\phi_h h^+} + A_{UU}^{\cos 2\phi_h h^-}$ black squares.
}
\end{figure}

Finally, we address the following question: can one determine, based on the available experimental data, whether the proposed relation in \eq{relation} between the $\cos \phi_h$ and $\cos 2\phi_h$ modulations holds?
It is difficult to answer this question using existing data. Indeed, $F_{UU}^{\cos 2\phi_h}$ receives both leading-power (LP) and NNLP contributions. As seen in Figs.~(\ref{fig:Acos2phi_HERMES}, \ref{fig:Acos2phi_COMPASS}), the LP contribution arising from the Boer–Mulders function convoluted with the Collins fragmentation function may not be dominant. Moreover, the Boer–Mulders contribution has opposite sign and approximately the same magnitude for $\pi^+$ and $\pi^-$ production (see \fig{Acos2phi_HERMES_LP}).
We exploit this approximate cancellation of the Boer–Mulders contribution in the sum of $\pi^+$ and $\pi^-$ and assume that the experimental yields of $\pi^+$ and $\pi^-$ are approximately equal. Using the experimental data, we construct the sum
\begin{align}
A_{UU}^{\cos 2\phi_h \pi^+} + A_{UU}^{\cos 2\phi_h \pi^-}
\end{align}
which approximately contains only NNLP terms.
which is expected to be dominated by NNLP terms. Similarly, for $F_{UU}^{\cos \phi_h}$, we form
\begin{align}
A_{UU}^{\cos \phi_h \pi^+} + A_{UU}^{\cos \phi_h \pi^-}
\end{align}
and compare $-q_T/Q,(A_{UU}^{\cos \phi_h,\pi^+} + A_{UU}^{\cos \phi_h,\pi^-})$ with $A_{UU}^{\cos 2\phi_h,\pi^+} + A_{UU}^{\cos 2\phi_h,\pi^-}$ in \fig{relation}.

For this comparison, we use the COMPASS data from Ref.~\cite{COMPASS:2014kcy}. Since the data are reported for unidentified charged hadrons, $h^\pm$, we assume dominance of charged pions.

If the relation were exact, the red circles would lie on top of the black squares in \fig{relation}. However, \fig{relation} shows that the relation in \eq{relation} is, at best, approximate: while the magnitudes and signs of the contributions are similar, their shapes differ. Note that this relation is expected to hold only at the level of $1/N_c \sim 30\%$ and in the region $q_T \ll Q$.

One should also exercise caution in interpreting \fig{relation}, as it does not represent a direct experimental measurement but rather an estimate based on several assumptions. A more detailed experimental study would be valuable for testing the validity of \eq{relation}. Based on the available evidence, we conclude that QCD dynamics in this regime is rich and that further theoretical and phenomenological investigations, both of our approximations and of the neglected contributions, are necessary.

\section{Conclusions and outlook}
\label{sec:conclusions}
In this paper, we utilized the rapidity factorization formalism~\cite{Balitsky:2017flc}, developed by one of the authors, in which next-to-next-to-leading-power ($\sim 1/Q^2$) corrections to the Drell–Yan process were derived in Refs.~\cite{Balitsky:2017gis,Balitsky:2021fer}. We extended this framework to the SIDIS process and derived NNLP corrections to the SIDIS hadronic tensor, including convolutions of unpolarized distributions, $f_1$, with unpolarized fragmentation functions, $D_1$, as well as Boer–Mulders functions, $h_1^\perp$, with Collins fragmentation functions, $H_1^\perp$. We obtained analytic expressions for the unpolarized SIDIS structure functions $F_{UU,T}$, $F_{UU,L}$, $F_{UU}^{\cos\phi_h}$, and $F_{UU}^{\cos2\phi_h}$. The structure functions $F_{UU,T}$ and $F_{UU}^{\cos2\phi_h}$ start at leading power and receive NNLP contributions; $F_{UU}^{\cos\phi_h}$ starts at next-to-leading power (NLP) and receives contributions at next-to-next-to-next-to-leading power (NNNLP), $1/Q^3$; while $F_{UU,L}$ starts at NNLP. We presented our results in both momentum and coordinate space, making the formulas suitable for phenomenological applications.

Using a simple model for TMD PDFs and TMD FFs, we provided numerical estimates and compared our results with data from the HERMES and COMPASS experiments. We also presented predictions for future measurements of $R_{\rm SIDIS}$ at Jefferson Lab and at the future Electron–Ion Collider. We compared our formulas with the existing literature and found overall agreement among different formalisms, although some differences remain. Further theoretical and phenomenological studies will be important to clarify these differences and to test the relations derived in this work. The size of the NLP and NNLP corrections indicates that they should be included in phenomenological analyses, especially for data at low $Q^2$. In this context, forthcoming measurements at Jefferson Lab and the future EIC will be particularly valuable when analyzed with NLP and NNLP corrections taken into account.

In future work, we plan to study polarized SIDIS structure functions and to investigate NNLP corrections arising from additional terms in the hadronic tensor that were not considered in this publication. We also aim to incorporate NLP and NNLP corrections into phenomenological studies with full TMD evolution.

\section*{Acknowledgments} 
We would like to thank Iain Stewart for correspondence and stimulating discussions, Leonard Gamberg and Alexey Vladimirov for discussions and comments on the manuscript.
This work was supported by the U.S. Department of Energy contract No.~DE-AC05-06OR23177, under which Jefferson Science Associates, LLC operates Jefferson Lab (I.B.,A.P.), and within the framework of the Saturated Glue (SURGE) Topical Theory
Collaboration (I.B.), and by the U.S. Department of Energy grant No.~DE-FG02-97ER41028 (I.B.), and by the National Science Foundation under Grants No.~PHY-2310031, No.~PHY-2335114 (A.P.). The research reported here is connected with the Quark-Gluon Tomography Collaboration supported by the U.S. Department of Energy, Office of Science,
Office of Nuclear Physics under contract DE-SC0023646 (A.P.).

\appendix

\section{Conversions}
\label{app:conv}

There exist several different notations for the vectors in hadron-hadron and in photon-hadron frames.
In order to facilitate the comparison of the results we will list the papers we investigate in our publication and comment on the notations.
\begin{itemize}
\item Pilo\~neta and Vladimirov in Ref.~\cite{Piloneta:2025jjb} explore both frames and provide final results in hadron-hadron frame. Transverse vectors in photon-hadron frame are denoted by $\perp$ while those in hadron-hadron frame are denoted by ${\bm T}$. The vectors in Eqs.~(B.3-B.6) of Ref.~\cite{Piloneta:2025jjb}are related to our notations as follows:
\begin{align}
\left[{\bm k}_1\right]_{\mbox{\tiny Ref.\cite{Piloneta:2025jjb}}} &= \left[{\bm k_\perp}\right]_{\rm our}\,, \\
\left[{\bm k}_2\right]_{\mbox{\tiny Ref.\cite{Piloneta:2025jjb}}} &= \left[-\frac{\bm p_\perp}{\zh}\right]_{\rm our}\,, \\
\left[{\bm h}_T\right]_{\mbox{\tiny Ref.\cite{Piloneta:2025jjb}}} &= \left[-{\hat {\bm h}}\right]_{\rm our}\,.
\end{align}

\item Ebert, Gao, and Stewart in Ref.~\cite{Ebert:2021jhy} work in hadron-hadron frame.
They parametrize $\qT = q_T (-1,0)$ and thus $\bfhp = (1,0)$.
In order to compare their notations for the convolutions we use 

\begin{align}
\left[{k}_{Tx}\right]_{\mbox{\tiny Ref.\cite{Ebert:2021jhy}}} &= \left[\bfhp \cdot {\bm k_T}\right]_{\rm our}= \left[\bfhp \cdot {\bm k_\perp}\right]_{\rm our}\,, \\
\left[{p}_{Tx}\right]_{\mbox{\tiny Ref.\cite{Ebert:2021jhy}}} &= \left[\bfhp \cdot {\bm p_T'}\right]_{\rm our}= \left[-\frac{\bfhp \cdot {\bm p_\perp}}{\zh}\right]_{\rm our}\,,
\end{align}

\item Bacchetta et al in Ref.~\cite{Bacchetta:2006tn} work in hadron-hadron frame.
The conversion of vectors read
\begin{align}
\left[{\bm p}_T\right]_{\mbox{\tiny Ref.\cite{Bacchetta:2006tn}}} &= \left[{\bm k_\perp}\right]_{\rm our}\,, \\
\left[{\bm k}_T\right]_{\mbox{\tiny Ref.\cite{Bacchetta:2006tn}}} &= \left[-\frac{\bm p_\perp}{\zh}\right]_{\rm our}\,, \\
\left[\hat {\bm h}\right]_{\mbox{\tiny Ref.\cite{Bacchetta:2006tn}}} &= \left[{\hat {\bm h}}\right]_{\rm our}\,.
\end{align}
Notice that in the convolution of Eq. (4.1) of Ref.~\cite{Bacchetta:2006tn} one should use $D^a(z,{\bm k_T^2})\to D^a(z,z^2 {\bm k_T^2})$.

\item Anselmino et al in Ref.~\cite{Anselmino:2011ch} work in photon-hadron frame.
The conversion of vectors read
\begin{align}
\left[{\bm k}_\perp\right]_{\mbox{\tiny Ref.\cite{Anselmino:2011ch}}} &= \left[{\bm k_\perp}\right]_{\rm our}\,, \\
\left[{\bm p}_\perp\right]_{\mbox{\tiny Ref.\cite{Anselmino:2011ch}}} &= \left[-\frac{\bm p_\perp}{\zh}\right]_{\rm our}\,, \\
\left[{\bm P_T}\right]_{\mbox{\tiny Ref.\cite{Anselmino:2011ch}}} &= \left[{ {\bm P_{h\perp}}}\right]_{\rm our}\,, \\
\left[\hat{\bm P}_T\right]_{\mbox{\tiny Ref.\cite{Anselmino:2011ch}}}&= \left[\hat {\bm h}\right]_{\rm our}\,.
\end{align}
\end{itemize}

\section{Structure functions from Refs.~\cite{Mulders:1995dh,Bacchetta:2006tn,Bacchetta:2008xw}}
\label{app:misc}
For completeness, in this Appendix we will report the structure functions  given by~\cite{Mulders:1995dh,Bacchetta:2006tn,Bacchetta:2008xw} using our notations~\footnote{In these formulas $h_1^\perp$ is for SIDIS.}
\begin{align} 
 F_{UU,T}	&= {\cal C} \bigl[ f_1 D_1 \bigr]
\,,\label{eq:FUU_Bacchetta}\\
F_{UU,L}	&= \frac{4 M^2}{Q^2}{\cal C} \biggl[  \frac{{\bm k_\perp^2}}{M^2} f_1 D_1 \biggr]
\,,\label{eq:FUUL_Bacchetta}\\
F_{UU}^{\cos\phi_h}
	&=
	\frac{2M_N}{Q}\,{\cal C}\biggl[
   	 \frac{\bfhp\cdot{\bm p_\perp}}{z \mh} 
	\biggl( x h\,H_{1}^{\perp }
   	+  \frac{\mh}{M}\,\,f_1 \frac{\tilde{D}^{\perp }}{z}\biggr)
	-  \frac{\bfhp\cdot{\bm k_\perp}}{M_N} \biggl( x  f^{\perp } D_1
   	+ \frac{\mh}{M_N}\,\,h_{1}^{\perp } \frac{\tilde{H}}{z}\biggr)\biggr] ,
	\label{eq:FUUcosphi_Bacchetta}\\
 F_{UU}^{\cos 2\phi_h} &= {\cal C}\left[\;\frac{2\, \bigl(\bfhp\cdot{\bm k_\perp}\bigr)\,
			\bigl(\bfhp\cdot{\bm p_\perp}\bigr)-{\bm k_\perp}\cdot{\bm p_\perp}}{zM\mh} h_1^{\perp} H_1^{\perp}\right] +
       \frac{4M^2}{Q^2} \; {\cal C}\left[\;\frac{2\,(\bfhp\cdot{\bm k_\perp})^2-{\bm k_\perp^2}}{2M^2}\,f_{1}\,D_{1}\;
	\right] 
\,.\label{eq:FUUcos2phi_Bacchetta}\end{align}
The equation of motion relations arise if one works in the tree level of hard interactions, see Ref.~\cite{Boussarie:2023izj}, and 
in the chiral even sector~\cite{Tangerman:1994bb, Mulders:1995dh, Bacchetta:2006tn} the equations read 
\begin{align} \label{eq:EOMrltn}
  x f^\perp &= x \tilde{f}^\perp + f_1 \,, \quad  x h = x \tilde{h} + \frac{k_\perp^2}{M^2}h_1^\perp,\\ \nn
  \frac{D^\perp}{z} &=  \frac{\tilde D^\perp}{z} + D_1
  \,, \quad \frac{H}{z}=
  \frac{\tilde H}{z} + \frac{p_\perp^2}{z^2M_h^2}H_1^\perp
  \,.
\end{align} 
Note that $k_\perp^2 = -{\bm k_\perp^2}$ and $p_\perp^2 =-{\bm p_\perp^2}$.
If one applies the Wandzura-Wilczek approximation, i.e. sets all dynamical functions labeled with a tilde to zero, then $F_{UU}^{\cos\phi_h}$ in Eq.~\eqref{eq:FUUcosphi_Bacchetta}, becomes, see e.g. Ref.~\cite{Boussarie:2023izj}:
\begin{align} \label{eq:FUUcosphiWW}
F_{UU}^{\cos\phi_h}
	&=
	\frac{2M}{Q}\,{\cal C}\biggl[
   	 -\frac{\bfhp\cdot{\bm p_\perp}}{z \mh} \frac{\bm k_\perp^2}{M}
	h_1^\perp H_{1}^{\perp }
	-  \frac{\bfhp\cdot{\bm k_\perp}}{M} f_1 D_1 \biggr] ,
\end{align}
and one can see that we reproduce our \eq{FUUcosphi-final}.
The next-to-leading power (NLP) terms present in $F_{UU,L}$, Eq.~\eqref{eq:FUUL}, and in the second term of $F_{UU}^{\cos 2\phi_h}$, Eq.~\eqref{eq:FUUcos2phi}, are derived in Bachetta et al Ref.~\cite{Bacchetta:2008xw} using parton model calculation of Anselmino et al, Ref.~\cite{Anselmino:2005nn}. The second term in Eq.~\eqref{eq:FUUcosphiWW} and Eq.~\eqref{eq:FUUcos2phi} is referred to as Cahn effect, Refs.~\cite{Cahn:1978se, Cahn:1989yf}.

One can also express convolutions in Eq.~\eqref{eq:FUU_Bacchetta},\eqref{eq:FUUL_Bacchetta},\eqref{eq:FUUcosphiWW}, and \eqref{eq:FUUcos2phi_Bacchetta} through Fourier transforms of products of TMDs in $b_T$ space~\cite{Boer:2011xd},
\begin{align}
\label{eq:FUU-bspace}
 {F}_{UU,T} &=   {\cal B}\left[\tilde f_{1}^{(0)}\, \tilde D_{1}^{(0)}\right]
\,,\\
\label{eq:FUUL-bspace}
 {F}_{UU,L} &= \frac{4}{Q^2} {\cal B}\left[\widetilde{k_\perp^2 f_{1}}^{(0)}\, \tilde D_{1}^{(0)}\right]
 \,,\\
\label{eq:FUUcosphi-bspace}
 F_{UU}^{\cos \phi_h}  &=  -\frac{2 M^2}{Q}{\cal B} \left[\tilde f_1^{(1)} \tilde D_1^{(0)} \right]- 
\frac{2 \mh}{Q M} {\cal B}\left[\widetilde{k_\perp^2 h_{1}^\perp}^{(0)}\, \tilde H_{1}^{\perp(1)}\right]
\,,\\
\label{eq:FUUcos2phi-bspace}
 F_{UU}^{\cos 2\phi_h}  &=  M\, \mh\; {\cal B}\left[\tilde h_{1}^{\perp (1)}\,\tilde H_{1}^{\perp (1)}\right]
+ \frac{M^4}{Q^2} \; {\cal B}[\tilde f_{1}^{(2)}\,\tilde D_{1}^{(0)}] \,.\end{align}

\providecommand{\href}[2]{#2}\begingroup\raggedright\endgroup

\end{document}